\documentclass[useAMS,fleqn,usenatbib]{mnras}

\usepackage{graphicx}
\usepackage{mathtools}
\usepackage[usenames,dvipsnames]{color}
\usepackage{amsmath,amssymb,latexsym}

\usepackage[T1]{fontenc} 
\usepackage{ae,aecompl}

\def\go{\mathrel{\raise.3ex\hbox{$>$}\mkern-14mu
             \lower0.6ex\hbox{$\sim$}}}
\def\lo{\mathrel{\raise.3ex\hbox{$<$}\mkern-14mu
             \lower0.6ex\hbox{$\sim$}}}

\newcommand{\be}{\begin{equation}}
\newcommand{\ee}{\end{equation}}
\newcommand{\bea}{\begin{eqnarray}}
\newcommand{\nn}{\nonumber}
\newcommand{\eea}{\end{eqnarray}}

\newcommand{\p}{\partial}

\title[A parameterised neutron star metric]{An accurate metric for the spacetime around neutron stars.}
\author[G.~Pappas]{George Pappas$^{1,2}$\thanks{E-mail:
gpappas@olemiss.edu}\\ 
$^{1}$Department of Physics and Astronomy, The University of Mississippi, University, MS 38677, USA\\
$^{2}$Departamento de F\'isica, CENTRA, Instituto Superior T\'ecnico, Universidade de Lisboa, Avenida Rovisco Pais 1, 
1049 Lisboa, Portugal}

\date{}

\pubyear{2016}

\begin{document}
\label{firstpage}
\pagerange{\pageref{firstpage}--\pageref{lastpage}}
\maketitle

\begin{abstract}
The problem of having an accurate description of the spacetime around neutron stars is of great astrophysical interest. For astrophysical applications, one needs to have a metric that captures all the properties of the spacetime around a neutron star. Furthermore, an accurate appropriately parameterised metric, i.e., a metric that is given in terms of parameters that are directly related to the physical structure of the neutron star, could be used to solve the inverse problem, which is to infer the properties of the structure of a neutron star from astrophysical observations. In this work we present such an approximate stationary and axisymmetric metric for the exterior of neutron stars, which is constructed using the Ernst formalism and is parameterised by the relativistic multipole moments of the central object. This metric is given in terms of an expansion on the Weyl-Papapetrou coordinates with the multipole moments as free parameters and is shown to be extremely accurate in capturing the physical properties of a neutron star spacetime as they are calculated numerically in general relativity. Because the metric is given in terms of an expansion, the expressions are much simpler and easier to implement, in contrast to previous approaches. For the parameterisation of the metric in general relativity, the recently discovered universal 3-hair relations are used to produce a 3-parameter metric. Finally, a straightforward extension of this metric is given for scalar-tensor theories with a massless scalar field, which also admit a formulation in terms of an Ernst potential. 
\end{abstract}


\section{Introduction}

Neutron stars (NSs) are excellent physics laboratories. Being astrophysical objects of high density and strong gravity, they can be used on the one hand to investigate the properties of matter at supranuclear densities and constrain the equation of state (EoS) and on the other hand to test the predictions of the established theory of gravity, general relativity (GR), as well as probe for possible modifications or deviations from it \citep{LattimerPrakash,Berti-Review2015}.    

The properties of isolated NSs are of relevance for systems such as low mass X-ray binaries (LMXBs) that might host them, were the interaction of the NS with its companion star result in astrophysical processes that can probe the spacetime around it, such as accretion discs or X-ray bursts. The NS properties are encoded in the surrounding spacetime and affect the motion of test particles, fluids or photons in the vicinity of the NS, encoding in turn these properties to the astrophysical observables.

Such astrophysical processes are for example the quasi-periodic oscillations (QPOs) of the X-ray flux observed from LMXBs \citep[for a review, see][]{lamb,derKlis} or the fluorescent iron lines also observed from these systems \citep[see for example][with exotic applications mostly]{IngramDone2012MNRAS,Bambi2013PhRvD,Bambi2015JCAP,Bambi2016arXiv}. In the former case the spacetime characteristic frequencies of orbital and precessional motion (which encode information of the structure of the NS) can be either straightforwardly related to the QPOs, as in the case of the relativistic precession model\footnote{Recently it was shown by \cite{Ingram2016MNRAS} that low frequency QPOs of thin accretion discs around black holes are related to the Lense-Thirring precession of the inner disc.} \citep[see][]{StellaVietri1998ApJ,StellaVietri1999PRL,stella} or the case of corrugation (c-)modes (and other modes) of thin accretion discs \citep[see for example][]{Tsang-Pappas2016ApJ}, or alternatively be more indirectly related as is the case of thick disc or donut-like models \citep[see for example][]{Rezzolla2003MNRAS}. In the latter case the information about the NS structure are encoded in both the orbital motion of the fluid in the accretion disc as well as in the orbits that the photons travel around the star.

Therefore, an accurate spacetime for the exterior of NSs that is also parameterised in terms of the properties of the structure of the central object would be very useful. A first approach on describing the spacetime around NSs was the slow rotation approximate solution by \cite{HT}, where the metric is expressed in terms of an expansion up to second order in the rotation (see for example \cite{berti-white}). Apart from this approach, there have been attempts to describe the spacetime exterior to NSs by analytic stationary and axisymmetric spacetimes that are not constrained to be only slowly rotating (see for example work by 
\cite{Stute,berti-stergioulas,Pappas2,Teich,Pachon,twosoliton,MankoRuiz2016}), where the spacetime is parameterised by a number of parameters and it is a vacuum solution of GR. Some of these analytic vacuum solutions have been found to be quite accurate (see for example \cite{Pappas2,twosoliton} for the two-soliton analytic solution of \cite{twosoliton2}), but a big obstacle in the implementation of these solutions to astrophysical problems has been the extremely complicated form that they have. Although the generating algorithm for stationary axisymmetric solutions in GR via the Ersnt potential \citep{ernst1} is very powerful (see \cite{sib1,SibManko,manko2,twosoliton2}) and can accommodate any number of parameters, the resulting metric expressions can be horrific. 

For this reason in this work we will take advantage of the powerful tool of the Ernst formalism and instead of constructing an exact vacuum solution we will construct an approximate vacuum solution in the form of an expansion in the Weyl-Papapetrou coordinates $(\rho,z)$ by starting with an expansion of the secondary Ernst potential $\xi$ which we will turn to an expansion of the Ernst potential $\mathcal{E}$. From this Ernst potential one can straightforwardly  calculate the metric functions of the \cite{Papapetrou} line element. The initial $\xi$ expansion will be expressed in terms of the relativistic multipole moments \citep{Geroch70I,Geroch70II,hansen}, therefore the resulting metric will be parameterised by the moments as well. 

This type of approximate spacetime solution can be extended from GR to the case of a scalar-tensor theory with a massless scalar field. The resulting scalar field and spacetime, expressed in the Jordan (physical) frame, will be parameterised by the multipole moments as they have been recently defined by \cite{PappasSTMoments} and by a set of coupling parameters characteristic to the specific theory, i.e., corresponding to a specific choice of the conformal factor that relates the Einstein and the Jordan frame and couples the scalar field to matter \citep[see for example][]{PappasSTRyan}.

\subsection{Executive summary}

This work is lengthy and there are some technical parts that may not be of interest to every reader.  
To help the reader navigate through these parts of the paper, we provide here a brief summary of the various topics. 

First we start by giving the main result of this paper, which is a stationary and axisymmetric spacetime parametrised in terms of the first five relativistic multipole moments, i.e., the mass $M$, the angular momentum $J$, the mass quadrupole $M_2$, the spin octupole $S_3$, and the mass hexadecapole $M_4$. The spacetime is given in the form of the \cite{Papapetrou} line elemet,
\be \label{Pap} ds^2=-f\left(dt-\omega d\varphi\right)^2+  f^{-1}\left[ e^{2\gamma} \left( d\rho^2+dz^2 \right)+ \rho^2 d\varphi^2 \right],
\ee
where ($\rho,z$) are the Weyl-Papapetrou coordinates and the metric functions $f,\;\omega,$ and $\gamma$ are given as 
\bea   f(\rho,z) \!\!\!\!\!\!&=&\!\!\!\!\!\!1-\frac{2 M}{\sqrt{\rho ^2+z^2}}+\frac{2 M^2}{\rho ^2+z^2}\nn\\
                                   &&\!\!\!\!\!\!+\frac{\left(M_2-M^3\right) \rho ^2-2 \left(M^3+M_2\right) z^2}{\left(\rho ^2+z^2\right)^{5/2}} \nn\\ 
                                   &&\!\!\!\!\!\!+\frac{2 z^2 \left(-J^2+M^4+2 M_2 M\right)-2 M M_2 \rho ^2}{\left(\rho ^2+z^2\right)^3}\nn\\
                                    &&\!\!\!\!\!\!+\frac{A(\rho,z)}{28 \left(\rho ^2+z^2\right)^{9/2}}+\frac{B(\rho,z)}{14 \left(\rho ^2+z^2\right)^5},\\
 \omega(\rho,z) \!\!\!\!\!\!&=&\!\!\!\!\!\!-\frac{2 J \rho ^2}{\left(\rho ^2+z^2\right)^{3/2}}   -\frac{2 J M \rho ^2}{\left(\rho ^2+z^2\right)^2}  +\frac{F(\rho,z)}{\left(\rho ^2+z^2\right)^{7/2}}\nn\\
                           && \!\!\!\!\!\!  +\frac{H(\rho,z)}{2 \left(\rho ^2+z^2\right)^4}+ \frac{G(\rho,z)}{4 \left(\rho ^2+z^2\right)^{11/2}} ,\\                             
               \gamma(\rho,z)\!\!\!\!\!\!&=&\!\!\!\!\!\!\frac{\rho ^2 \left(J^2 \left(\rho ^2-8 z^2\right)+M \left(M^3+3 M_2\right) \left(\rho ^2-4 z^2\right)\right)}{4 \left(\rho
   ^2+z^2\right)^4}\nn\\
           &&\!\!\!\!\!\!-\frac{M^2 \rho ^2}{2 \left(\rho ^2+z^2\right)^2},                     
                                    \eea
where,
\bea   A(\rho,z) \!\!\!\!\!\!&=& \!\!\!\!\!\! \left[8 \rho ^2 z^2 \left(24 J^2 M+17 M^2 M_2+21 M_4\right)\right.\nn\\
      &&\!\!\!\!\!\! +\rho ^4 \left(-10 J^2 M+7 M^5+32 M_2 M^2-21 M_4\right)\nn\\
      &&\!\!\!\!\!\!\left.+8 z^4 \left(20 J^2 M-7 M^5-22 M_2 M^2-7 M_4\right)\right] , \\
   B(\rho,z)  \!\!\!\!\!\!&=&\!\!\!\!\!\!  \left[\rho ^4 \left(10 J^2 M^2+10 M_2 M^3+21 M_4M+7 M_2^2\right) \right.\nn\\.
          &&\!\!\!\!\!\! +4 z^4 \left(-40 J^2 M^2-14 J S_3+7 M^6+30 M_2 M^3 \right.\nn\\
          &&\!\!\!\!\!\!\left.+14 M_4 M+7 M_2^2\right)-4 \rho ^2 z^2 \left(27 J^2 M^2-21 J S_3\right.\nn\\
   &&\!\!\!\!\!\!\left.\left.+7 M^6+48 M_2 M^3+42 M_4 M+7 M_2^2\right)\right],\\
 H(\rho,z) \!\!\!\!\!\!&=&\!\!\!\!\!\!  \left[4 \rho ^2 z^2 \left(J \left(M_2-2 M^3\right)-3 M S_3\right)\right.\nn\\
                                    &&\!\!\!\!\!\!\left.+\rho ^4 \left(J M_2+3 M S_3\right)\right] \\
   G(\rho,z) \!\!\!\!\!\!&=&\!\!\!\!\!\!\left[\rho ^2 \left(J^3 \left(-\left(\rho ^4+8 z^4-12 \rho ^2 z^2\right)\right)\right.\right.\nn\\
                               &&\!\!\!\!\!\!+J M \left(\left(M^3+2 M_2\right) \rho ^4-8 \left(3 M^3+2 M_2\right) z^4\right.\nn\\
                               &&\!\!\!\!\!\!\left.+4 \left(M^3+10 M_2\right) \rho ^2 z^2\right)\nn\\
                               &&\!\!\!\!\!\!\left.\left.+M^2 S_3 \left(3 \rho ^4-40 z^4+12 \rho ^2 z^2\right)\right)\right] \\
   F(\rho,z) \!\!\!\!\!\!&=&\!\!\!\!\!\! \left[\rho ^4 \left(S_3-J M^2\right)-4 \rho ^2 z^2 \left(J  M^2+S_3\right)\right] .   \eea
The derivation of this spacetime will be discussed in section 2 where we give some background on finding analytic solutions of the Einstein field equations in GR and setup the setting for constructing an approximate spacetime for the exterior of NSs. In order to customise this general spacetime to the case of NSs, we make use of the recently found properties of the multipole moments of NS spacetimes \citep{Pappas:2013naa,Stein2014ApJ,YagietalM4}. This also provides a more economic description of NS spacetimes in terms of the number of parameters, something that can facilitate the astrophysical application of the spacetime \citep[see for example][]{Pappas2015Unified}.    
We further explore some of the general properties of this spacetime in Section 3. In Section 4 we test how well the present solution compares to the \cite{HT} spacetime, the two-soliton analytic spacetime, and numerically calculated spacetimes and discuss the range of parameters for which this spacetime is appropriate. In Section 5 we construct the corresponding scalar-tensor solution by extending the aforementioned GR solution to the case of scalar-tensor theory with a massless scalar field. Finally we end in Section 6 with the conclusions. 
Throughout we use geometric units, where $G=c=1$ and the masses are given in $km$, unless some other unit is specified.

\section{The approximate NS spacetime}
\label{sec:2}

One can calculate a spacetime for a NS by implementing the following two approaches, either use a slow rotation limit and have a solution as it is given by the \cite{HT} approach \citep[see also][]{berti-white}, or implement a numerical algorithms for solving the full Einstein field equations for axisymmetric spacetimes around rotating fluid configurations (for example see \cite{Sterg} and for more details the review by \cite{Lrr}). Alternatively, a lot of work has been done on analytic axisymmetric spacetimes that can match the exterior of NSs, as was mentioned in the Introduction. 
These last attempts are based on algorithmically constructing parameterised stationary and axisymmetric solutions of the vacuum Einstein's field equations \citep[see][]{sib1,SibManko,manko2,twosoliton2}.  The approximate solution that is presented here will not make use of these algorithms and will only be based on solving the Ernst equations.   

\subsection{General setup}

The vacuum region of a stationary and axially symmetric spacetime in GR, i.e., a spacetime that admits a timelike Killing vector associated to time translations and a spacelike Killing vector associated to rotations around an axis of symmetry, can be described, as we have mentioned, by the line element (\ref{Pap}) introduced by \cite{Papapetrou}. 
In the line element (\ref{Pap}), $f,\;\omega,$ and $\gamma$ are functions of the Weyl-Papapetrou coordinates ($\rho,z$). By introducing the complex potential $\mathcal{E}(\rho,z)=f(\rho,z)+\imath\psi(\rho,z)$, where $\psi$ is the scalar twist of the timelike Killing vector, \cite{ernst1} reformulated the Einstein field equations in the form of the complex equation
\be \label{ErnstE}
(Re(\mathcal{E}))\nabla^2\mathcal{E}=\nabla\mathcal{E}\cdot\nabla\mathcal{E},
\ee
where $\nabla$ and $\nabla^2$ are respectively the gradient and the Laplacian in flat cylindrical coordinates $(\rho,z,\varphi)$, while the third metric function satisfies the following set of equations
\bea \frac{\p\gamma}{\p \rho}&=&\frac{\rho}{4f^2}\left[\left(\frac{\p f}{\p\rho}\right)^2-\left(\frac{\p f}{\p z}\right)^2\right]\nn\\
&&-\frac{f^2}{4\rho}\left[\left(\frac{\p \omega}{\p\rho}\right)^2-\left(\frac{\p \omega}{\p z}\right)^2\right], \label{gammar}\\
\frac{\p\gamma}{\p z}&=&\frac{1}{2}\left[\frac{\rho}{f^2}\frac{\p f}{\p\rho}\frac{\p f}{\p z}-\frac{f^2}{\rho}\frac{\p \omega}{\p\rho}\frac{\p \omega}{\p z}\right]. \label{gammaz}
\eea
The scalar twist $\psi(\rho,z)$ is related to the metric function $\omega(\rho,z)$ through the identity \citep{ernst1}
\be f^{-2}\nabla \psi=-\rho^{-1}\hat{n}  \times\nabla\omega, \label{omegaID}\ee
where as before $\nabla$ is the gradient in the cylindrical flat coordinates and $\hat{n}$ is a unit vector in the azimuthal direction. 

One can generate a solution of the Ernst equation by making a choice for the Ernst potential along the axis of symmetry of the spacetime, in the form of a rational function
\be
\mathcal{E}(\rho=0,z)=e(z)=\frac{P(z)}{R(z)},
\label{Eernst}
\ee
where $P(z), R(z)$ are polynomials of $z$ of order $n$ with complex coefficients in general. A choice of these polynomials introduces a number of parameters in the form of the coefficients of the polynomials, which can later be given a specific physical meaning. An example of this form of solutions is the vacuum two-soliton solution (proposed by \cite{twosoliton2}) that is generated from the ansatz 
\be
\label{2soliton}
e(z)=\frac{(z-M-ia)(z+ib)-k}{(z+M-ia)(z+ib)-k},
\ee
where all the parameters $M, a, k, b$ are real (for more details on the algorithm for generating solutions and for the two-soliton spacetime in particular see \cite{twosoliton2,twosoliton,MankoRuiz2016}).
The parameters that are introduced by this ansatz can be related to physical properties of the spacetime and the NS and in particular they can be related to the relativistic multipole moments. The first few multipole moments of the two-soliton spacetime as they are expressed in terms of the four parameters introduced are,
\bea
\label{moments2}
M_0&=&M,\quad  M_2=-(a^2-k)M,\nn\\
 M_4&=&\left[ a^4 - (3a^2 -2ab + b^2)k + k^2 - \frac{1}{7}kM^2\right] M\nn\\
S_1&=& aM ,\quad S_3=- [a^3 -(2a - b)k]M,
\eea
where $M_0=M$ is the mass, $M_2=Q$ is the quadrupole moment, $M_4$ is the mass hexadecapole, $S_1=J$ is the angular momentum, and $S_3$ is the spin octupole moment. 

In principle one could introduce a very large number of parameters increasing this way the accuracy of the matching between an actual NS spacetime and the analytic solution (see for example \cite{Teich}), but as one increases the order of the polynomials of the axis ansatz, one also dramatically increases the complexity of the analytic solution (see for example how the \cite{Pachon} solution, which is a third order ansatz, compares to the \cite{twosoliton2} solution, which is a second order ansatz, even though the former ansatz has been constrained to have the same number of parameters as the latter ansatz).  

To avoid this complication, we will try to generate an approximate spacetime that is based on an expansion of the Ernst potential in terms of the Weyl-Papapetrou coordinates. \cite{fodor:2252} have shown that the secondary Ernst potential, 
\be \xi=\frac{1-\mathcal{E}}{1+\mathcal{E}}\,, \ee
admits an expansion in terms of the Weyl-Papapetrou coordinates that is of the form,
\be \tilde{\xi}=(1/\bar{r})\xi=\sum_{i,j=0}^{\infty} a_{ij}\bar{\rho}^i\bar{z}^j, \ee
where $\bar{\rho}=\rho/(\rho^2+z^2)$, $\bar{z}=z/(\rho^2+z^2)$, and $\bar{r}^2=\bar{\rho}^2+\bar{z}^2=r^{-2}=(\rho^2+z^2)^{-1}$ are asymptotic coordinates related to the usual Weyl-Papapetrou coordinates, while the complex coefficients $a_{ij}$ are constrained to satisfy specific relations between them (due to the field equations) and are related to the relativistic multipole moments of the spacetime \citep{fodor:2252}. 

The shot for approximate solution will be constructed as follows. We will first truncate the expansion of $\tilde{\xi}$, which will result in an expansion for $\xi$ of the form
\be \xi=\frac{1}{\sqrt{\rho^2+z^2}}\sum_{i,j=0}^{n,k} a_{ij}\left(\frac{\rho}{\rho^2+z^2}\right)^i\left(\frac{z}{\rho^2+z^2}\right)^j. \label{xi}\ee
From this, we can calculate the Ernst potential 
\be \mathcal{E}=\frac{1-\xi(\rho,z)}{1+\xi(\rho,z)}. \label{EofXi}\ee
This admits an infinite expansion in powers of $\bar{r}$, so we will truncate it again up to the required order. The real part of $\mathcal{E}$ is the metric function $f(\rho,z)$, while the imaginary part is the function $\psi(\rho,z)$, from which the metric function $\omega(\rho,z)$ can be calculated through the identity (\ref{omegaID}). 
 This equation can be integrated to give the metric function $\omega(\rho,z)$ in terms of the moments \citep[see discussion by][]{Ryan95}. The last remaining metric function, $\gamma(\rho,z)$ can be calculated by integrating equations (\ref{gammar},\ref{gammaz}).

\subsection{The metric in terms of moments}

As it was outlined in the previous subsection, we will start with the expansion of $\xi$, up to some order, given by equation (\ref{xi}). The coefficients $a_{ij}$ in that expression can be given in terms of the coefficients $a_{0j}=m_j$ of the expansion of $\tilde{\xi}$ along the axis of symmetry, $\tilde{\xi}(\bar{\rho}=0)=\sum_{j=0}^{\infty} m_j\bar{z}^j$. Also $a_{ij}=0$ if $i$ is an odd number \cite[see][for details]{fodor:2252}. The coefficients $m_j$ are related to the relativistic multipole moments of the spacetime and can be found by solving the expressions given by \cite{fodor:2252}. In Appendix  \ref{app:A} we give the relations between $a_{ij}$, $m_j$, and relativistic multipole moments.

We will start by setting the order of the expansion of the potential $\xi$ to be $\mathcal{O}(\bar{r}^6)$, which means that the sum in equation (\ref{xi}) will have terms of order $\mathcal{O}(\bar{r}^5)$ and the coefficients $a_{ij}$ that will appear in the expansion will have to satisfy $i+j\leq5$, since any higher summation term will correspond to a higher order term. Also, these coefficients will introduce terms of multipole order up to $S_5$. If we further require the same order expansion for the Ernst potential $\mathcal{E}$, then no term with a coefficient $a_{ij}$ with $i+j>5$ could appear. Therefore, the potential $\xi$ will have the expansion

\begin{align} \xi(\rho,z)&=\frac{1}{r}\left(a_ {00} + \frac {z a_ {01}} {\rho^2 + z^2} + \frac {z^2 a_ {02}} {\left (\rho^2 + z^2 \right)^2} + \frac {z^3 a_ {03}} {\left (\rho^2 + z^2 \right)^3} \right.\nn\\
                      &+ \frac {z^4 a_ {04}} {\left (\rho^2 + z^2 \right)^4}+ \frac {z^5 a_ {05}} {\left (\rho^2 + z^2 \right)^5} 
+ \frac {\rho^2 a_ {20}} {\left (\rho^2 + z^2 \right)^2} + \frac {\rho^2 z a_ {21}} {\left (\rho^2 + z^2 \right)^3}\nn\\
                     & \left.+ \frac {\rho^2 z^2 a_ {22}} {\left (\rho^2 + z^2 \right)^4} + \frac {\rho^2 z^3 a_ {23}} {\left (\rho^2 + z^2 \right)^5} 
                     + \frac {\rho^4 a_ {40}} {\left (\rho^2 + z^2 \right)^4} + \frac {\rho^4 z a_ {41}} {\left (\rho^2 + z^2 \right)^5} 
                     \right), \label{xi2}
            \end{align}
where, 
as we have mentioned earlier, the coefficients $a_{ij}$ are complex numbers which can be expressed in terms of the coefficients $m_j=a_{0j}$. With $\xi$ at hand it is straightforward to calculate the Ernst potential $\mathcal{E}$ from equation (\ref{EofXi}). The metric function $f(\rho,z)$ can then be evaluated to be the real part of the Ernst potential, while the function $\psi(\rho,z)$ will be the imaginary part,
\bea f(\rho,z)&=&\frac{1}{2}\left(\mathcal{E}+\mathcal{E}^*\right),\\
         \psi(\rho,z)&=&\frac{1}{2 i}\left(\mathcal{E}-\mathcal{E}^*\right). 
\eea
We should also note here that under the assumption of equatorial reflection symmetry, a reasonable assumption for rotating fluid configurations such as NSs, the coefficients $a_{ij}$ are real for even $j$ and imaginary for odd $j$. Using the relations between the $a_{ij}$ coefficients and the relations for the relativistic multipole moments, the metric function $f(\rho,z)$ will take the form,   
\bea   f(\rho,z) \!\!\!\!\!\!&=&\!\!\!\!\!\!1-\frac{2 M}{\sqrt{\rho ^2+z^2}}+\frac{2 M^2}{\rho ^2+z^2}\nn\\
                                   &&\!\!\!\!\!\!+\frac{\left(M_2-M^3\right) \rho ^2-2 \left(M^3+M_2\right) z^2}{\left(\rho ^2+z^2\right)^{5/2}} \nn\\ 
                                   &&\!\!\!\!\!\!+\frac{2 z^2 \left(-J^2+M^4+2 M_2 M\right)-2 M M_2 \rho ^2}{\left(\rho ^2+z^2\right)^3}\nn\\
                                    &&\!\!\!\!\!\!+\frac{A(\rho,z)}{28 \left(\rho ^2+z^2\right)^{9/2}}+\frac{B(\rho,z)}{14 \left(\rho ^2+z^2\right)^5},\nn
                                    \eea
where,
\bea   A(\rho,z) \!\!\!\!\!\!&=& \!\!\!\!\!\! \left[8 \rho ^2 z^2 \left(24 J^2 M+17 M^2 M_2+21 M_4\right)\right.\nn\\
      &&\!\!\!\!\!\! +\rho ^4 \left(-10 J^2 M+7 M^5+32 M_2 M^2-21 M_4\right)\nn\\
      &&\!\!\!\!\!\!\left.+8 z^4 \left(20 J^2 M-7 M^5-22 M_2 M^2-7 M_4\right)\right] , \nn\\
   B(\rho,z)  \!\!\!\!\!\!&=&\!\!\!\!\!\!  \left[\rho ^4 \left(10 J^2 M^2+10 M_2 M^3+21 M_4M+7 M_2^2\right) \right.\nn\\.
          &&\!\!\!\!\!\! +4 z^4 \left(-40 J^2 M^2-14 J S_3+7 M^6+30 M_2 M^3 \right.\nn\\
          &&\!\!\!\!\!\!\left.+14 M_4 M+7 M_2^2\right)-4 \rho ^2 z^2 \left(27 J^2 M^2-21 J S_3\right.\nn\\
   &&\!\!\!\!\!\!\left.\left.+7 M^6+48 M_2 M^3+42 M_4 M+7 M_2^2\right)\right].\nn
   \eea
where, in order to calculate the metric function in the form of an expansion, we have expanded the Ernst potential up to order $\mathcal{O}(\bar{r}^6)$, as we had done for $\xi$.\footnote{We should note that up to the chosen order, the expansion has $a_{05}$ terms as well, which would introduce an $S_5$ dependence. Since we are interested in parameters only up to $M_4$ we have set the $a_{05}$ terms to zero.} 
A similar expression can be derived for $\psi(\rho,z)$, which we will not reproduce here, since the object of interest is the metric function $\omega(\rho,z)$ instead.   

The function $\omega(\rho,z)$ can be evaluated from the function $\psi(\rho,z)$ by using the identity (\ref{omegaID}),\footnote{The identity results in a pair of equations which give the derivatives of $\omega$ in terms of $f$ and the derivatives of $\psi$. These expressions are also expanded up to $\mathcal{O}(\bar{r}^6)$, so that they can be integrated.} which after integration gives

 \bea  \omega(\rho,z) \!\!\!\!\!\!&=&\!\!\!\!\!\!-\frac{2 J \rho ^2}{\left(\rho ^2+z^2\right)^{3/2}}   -\frac{2 J M \rho ^2}{\left(\rho ^2+z^2\right)^2}  
                                                                 +\frac{F(\rho,z)}{\left(\rho ^2+z^2\right)^{7/2}}\nn\\
                           && \!\!\!\!\!\!  +\frac{H(\rho,z)}{2 \left(\rho ^2+z^2\right)^4}+ \frac{G(\rho,z)}{4 \left(\rho ^2+z^2\right)^{11/2}} ,\nn\eea
 where,
 \bea             H(\rho,z) \!\!\!\!\!\!&=&\!\!\!\!\!\!  \left[4 \rho ^2 z^2 \left(J \left(M_2-2 M^3\right)-3 M S_3\right)\right.\nn\\
                                    &&\!\!\!\!\!\!\left.+\rho ^4 \left(J M_2+3 M S_3\right)\right] \nn\\
   G(\rho,z) \!\!\!\!\!\!&=&\!\!\!\!\!\!\left[\rho ^2 \left(J^3 \left(-\left(\rho ^4+8 z^4-12 \rho ^2 z^2\right)\right)\right.\right.\nn\\
                               &&\!\!\!\!\!\!+J M \left(\left(M^3+2 M_2\right) \rho ^4-8 \left(3 M^3+2 M_2\right) z^4\right.\nn\\
                               &&\!\!\!\!\!\!\left.+4 \left(M^3+10 M_2\right) \rho ^2 z^2\right)\nn\\
                               &&\!\!\!\!\!\!\left.\left.+M^2 S_3 \left(3 \rho ^4-40 z^4+12 \rho ^2 z^2\right)\right)\right] \nn\\
   F(\rho,z) \!\!\!\!\!\!&=&\!\!\!\!\!\! \left[\rho ^4 \left(S_3-J M^2\right)-4 \rho ^2 z^2 \left(J  M^2+S_3\right)\right] \nn
   \eea

 The remaining undetermined metric function $\gamma(\rho,z)$ can be calculated, in a similar manner from the expressions (\ref{gammar},\ref{gammaz}), to be
 
 \bea \gamma(\rho,z)\!\!\!\!\!\!&=&\!\!\!\!\!\!\frac{\rho ^2 \left(J^2 \left(\rho ^2-8 z^2\right)+M \left(M^3+3 M_2\right) \left(\rho ^2-4 z^2\right)\right)}{4 \left(\rho
   ^2+z^2\right)^4}\nn\\
           &&\!\!\!\!\!\!-\frac{M^2 \rho ^2}{2 \left(\rho ^2+z^2\right)^2}. \nn \eea
 Thus, with the three functions $f(\rho,z)$, $\omega(\rho,z)$, and $\gamma(\rho,z)$ we have fully determined the spacetime, which is expressed in terms of the relativistic multipole moments, and is given by the line element (\ref{Pap}). For this metric, it can be verified that the spacetime is Ricci flat up to order $\mathcal{O}(\bar{r}^6)$, i.e., 
 \be R_{ab}=0+\mathcal{O}(\bar{r}^6).\nn\ee
 
\subsection{The NS spacetime}
 
 The metric as it is given so far can describe up to the order of $\mathcal{O}(\bar{r}^6)$ the vacuum spacetime around any object with a specific set of relativistic multipole moments up to $M_4$. Therefore, in principle, if the appropriate multipole moments were used this metric could approximate the spacetime around a black hole, a NS, a quark star or any other object that has no other extra fields.\footnote{The same construction could be done with the addition of electromagnetic multipole moments, since electro-vacuum spacetimes also admit a description in terms of an Ernst potential and the definition of multipole moments extends to these cases as well \citep[see][]{ernst2,Hoenselaers1990,SotiriouMoments,stephani}}  
 
 To describe a NS one needs to prescribe the right set of multipole moments. Recent work by \cite{Pappas:2013naa,YagietalM4} has shown that  
 for NSs (and this has been extended to quark stars as well by \cite{YagietalM4}) the first few relativistic multipole moments can be expressed in a Kerr-like fashion as,
 \bea  M_2 &=& -\alpha j^2 M^3 ,\\
          S_3 &=& -\beta j^3  M^4, \\
          M_4 &=& \; \;\; \gamma j^4 M^5, \\
              &&\ldots ,\nn
     \eea
where $M$ is the mass and $j$ is the spin parameter, i.e., the angular momentum over the mass $J/M^2$. The difference from the Kerr case is that for NSs the coefficients $\alpha$, $\beta$, and $\gamma$ are not equal to $1$ but instead can be much larger \citep{poisson,pappas-apostolatos,PappasMoments2,Pappas:2013naa,YagietalM4}. Furthermore it has been shown by \cite{Pappas:2013naa,Stein2014ApJ,YagietalM4} that for realistic EoSs the NS multipole moments are not all independent between them. In particular one can express the first moments that are higher than the quadrupole in terms of the quadrupole itself. Specifically, if we define the reduced moments as
\be \bar{M}_n=\frac{M_n}{j^nM^{n+1}}, \; \bar{S}_n=\frac{S_n}{j^nM^{n+1}}, \ee
then the spin octupole and the mass hexadecapole of a NS will be related to the quadrupole by relations of the form 
\be \label{momentsFit}  y=A+B_1 x^{\nu_1}+B_2 x^{\nu_2} ,\ee 
where $y$ can be either $\sqrt[3]{-\bar{S}_3}=\sqrt[3]{\beta}$ or $\sqrt[4]{\bar{M}_4}=\sqrt[4]{\gamma}$ and $x$ is $\sqrt{-\bar{M}_2}=\sqrt{\alpha}$. Therefore the first higher moments of a NS spacetime can be expressed in terms of only three parameters, the mass $M$, the angular momentum $J$, and the quadrupole $M_2=Q$. 
%
%
%
Additionally, these relations between the moments have been found to be approximately independent of the realistic EoS used, which means that if one were to use these expressions to produce a spacetime, then one could have an EoS independent description of the spacetime around a NS. To clarify the last statement, using the universal relations between the moments, one could construct a spacetime metric parameterised by only three parameters (mass, spin, and quadrupole) which will be suitable to describe the exterior of any NS constructed using any of the realistic EoSs that we have developed so far. 

For the construction of the spacetime in the previous subsection we used the first five moments, $M,J,M_2,S_3$ and $M_4$, therefore we will need the relation between the quadrupole and $S_3$ \citep{Pappas:2013naa} as well as the relation between the quadrupole and $M_4$ \citep{YagietalM4}, i.e., 
\bea y_1&=&-0.36+1.48\, x^{0.65},\label{uniRelS3}\\
         y_2&=&-4.749+0.27613\, x^{1.5146}+5.5168\, x^{0.22229},\label{uniRelM4}\eea
where $y_1=\sqrt[3]{-\bar{S}_3}=\sqrt[3]{\beta}$, $y_2=\sqrt[4]{\bar{M}_4}=\sqrt[4]{\gamma}$ and $x=\sqrt{-\bar{M}_2}=\sqrt{\alpha}$. We should note here that these expressions do not come from the Table I of \cite{YagietalM4}, because those fits were constructed for both NSs and quark stars. Instead the expressions for $y_1$ and $y_2$ come from \cite{Pappas:2013naa} and only the NS data of \cite{YagietalM4} (see Appendix \ref{app:C}) respectively. Therefore, the description of the spacetime and the various properties of the geodesics that we will calculate, will depend on the dimensional mass $M$, expressed in units of km, and two additional dimensionless parameters, i.e., the spin parameter $j=J/M^2$ and the reduced quadrupole $\alpha=-M_2/(j^2M^3)$. 

%
%

\subsection{Quasi-isotropic coordinates}
\label{sec:QuasIsoCoord}

When constructing numerical NS models with numerical schemes like the RNS code \citep{Sterg} the coordinate system usually used is that of quasi-isotropic coordinates. In these coordinates the metric is expressed in the form,
\begin{align}  ds^2&=-e^{2\nu} dt^2 + r^2\sin^2\theta B^2 e^{-2\nu} (d\varphi-\omega dt)^2 + e^{2\alpha} (dr^2 + r^2 d\theta^2) \nn\\
                             &=g_{tt}dt^2 + 2 g_{t\varphi} dtd\varphi +g_{\varphi\varphi}d\varphi^2 + g_{rr} (dr^2 + r^2 d\theta^2).
\end{align} 
Therefore one might be interested in the transformation from Weyl-Papapetrou $(\rho,z)$ coordinates to quasi-isotropic $(r,\theta)$ coordinates. The Weyl-Papapetrou coordinate $\rho$ is defined from the determinant of the $t-\varphi$ part of the metric as $\rho^2=g_{t\varphi}^2-g_{tt}g_{\varphi\varphi}$. It is therefore straightforward to show that the coordinate $\rho$ is given in terms of $r$ and $\theta$ as 
\be \rho(r,\mu) = r\sin\theta B(r,\theta) = r \sqrt{1-\mu^2} B(r,\mu), \ee
where $\mu=\cos\theta$. In vacuum, the coordinate $\rho$ is a harmonic function and the coordinate $z$ is its harmonic conjugate. From this property one can derive the form of the coordinate $z$ as a function of $r$ and $\theta$ \citep[see][for details]{Pappas2008CQG}, which is given from the expression,
\be z(r,\mu)= \int_0^{\mu} \left(r^2\frac{\p B}{\p r}+rB \right)d\mu' ,\ee
where the integration takes place in vacuum at constant radius $r$ and from the equatorial plane $\mu=0$ up to the required angle $\mu$.\footnote{Of course since $z$ is a harmonic function the path of integration doesn't matter as long as we are in vacuum, but the expression given here is not the general one. Instead it is a special case that assumes integrations along constant $r$.} As one can see, the coordinate transformation depends on knowing the function $B$ of the quasi-isotropic metric. As it was shown by \cite{BI}, the metric function $B$ has the form,
\begin{align} 
B &= \sum_{l=0}^{\infty}B_{2l}(r)T_{2l}^{1/2}(\mu)\nn\\
      &=  \left(\frac{\pi}{2}\right)^{1/2}\left(1+\frac{\tilde{B}_0}{r^2}\right)T_0^{1/2}(\mu)
                 +\left(\frac{\pi}{2}\right)^{1/2}\frac{\tilde{B}_2}{r^4}T_2^{1/2}(\mu)+\ldots\nn\\
       &= 1+\frac{\tilde{B}_0}{r^2} +(4\mu^2-1)\frac{\tilde{B}_2}{r^4}+\ldots,           \label{third}
\end{align}
where the functions $T_{2l}^{1/2}(\mu)$ are the Gegenbauer polynomials \citep[defined in][]{BI} and the coefficients $\tilde{B}_{2l}$ depend on the internal structure of the particular NS. From numerical models one can see that the general behaviour for the zero order coefficient is $\tilde{B}_0/M^2\equiv b_0 =-1/4+ \tilde{\beta}_0 j^2$ \citep{pappas-apostolatos,Morsink2014}, while for the next order coefficient the behaviour is $\tilde{B}_2/M^4\equiv b_2 = \tilde{\beta}_2 j^2$. One can also see that $\tilde{\beta}_0$ can be expressed as a function of the reduced quadrupole $\alpha$ of the form, $\tilde{\beta}_0\propto \alpha^{-\nu}$, where $\nu$ is positive, while $\tilde{\beta}_2$ can be related to $\alpha$ linearly, i.e.,  $\tilde{\beta}_2\propto \alpha$. The coefficients $\tilde{\beta}_0$ and $\tilde{\beta}_2$ seem to have a weak dependence on the spin parameter $j$ and their dependence on $\alpha$ seems to be approximately EoS universal, but the further exploration of this property is beyond the scope of this work.\footnote{The coefficients $\tilde{B}_0$ and $\tilde{B}_2$ as well as the moments up to $M_4$ have been calculated for a large number of models and for a variety of EoSs in the context of the work by \cite{YagietalM4}.} 
 
Finally, in order to have the metric in the quasi-isotropic form one only needs the transformation of the $r$-$\theta$ part of the metric, which is,
\be \frac{e^{2\gamma}}{f}(d\rho^2+dz^2)=\frac{e^{2\gamma}}{f}\left[\left(\frac{\p\rho}{\p \varpi}\right)^2+\left(\frac{\p\rho}{\p \zeta}\right)^2\right](dr^2+r^2d\theta^2), \ee
where $\varpi=r\sin\theta$ and $\zeta=r\cos\theta$.

\section{Properties of the spacetime}
\label{sec:observables}

\begin{figure*}
\centering
\includegraphics[width=.25\textwidth]{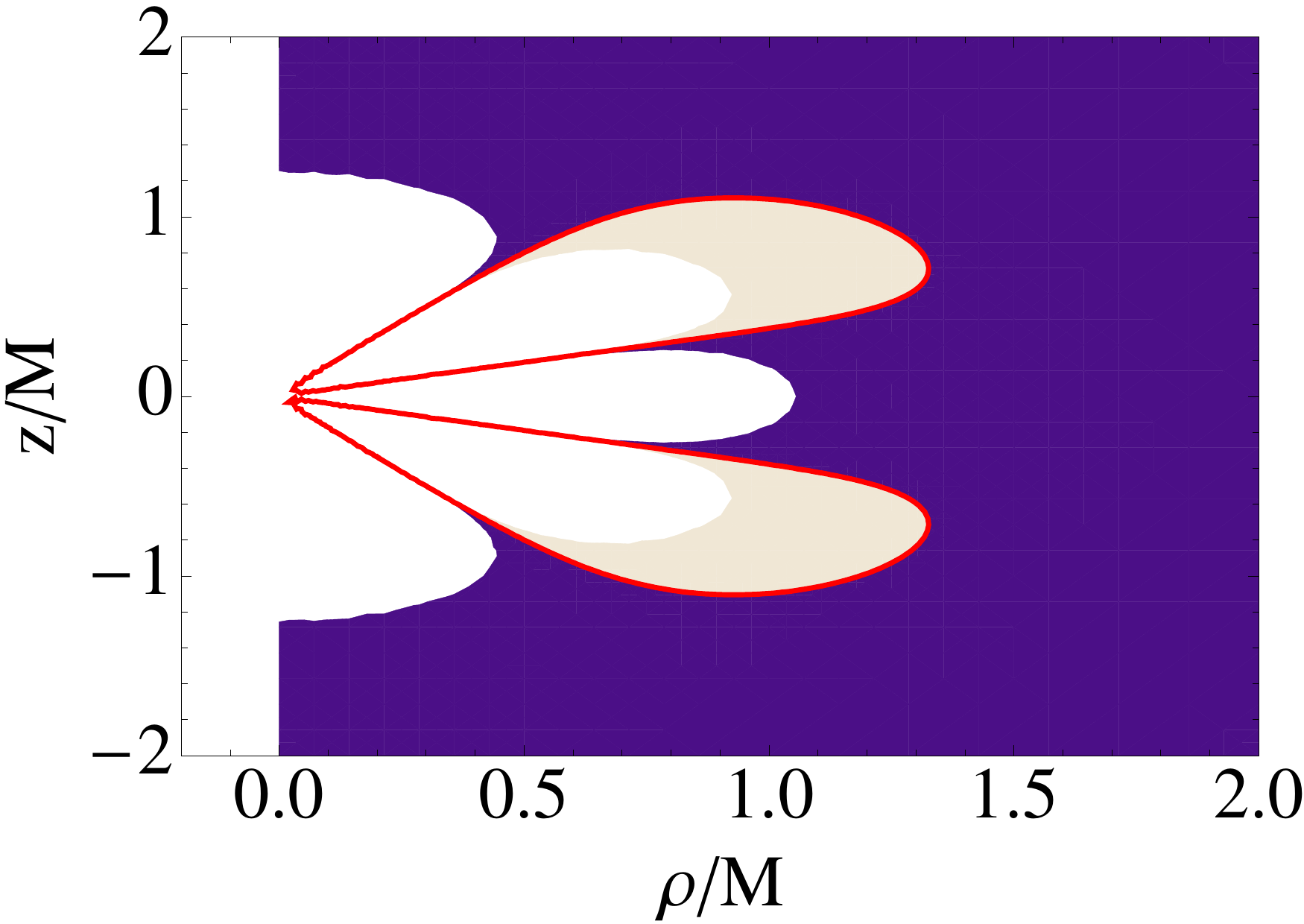} 
\includegraphics[width=.25\textwidth]{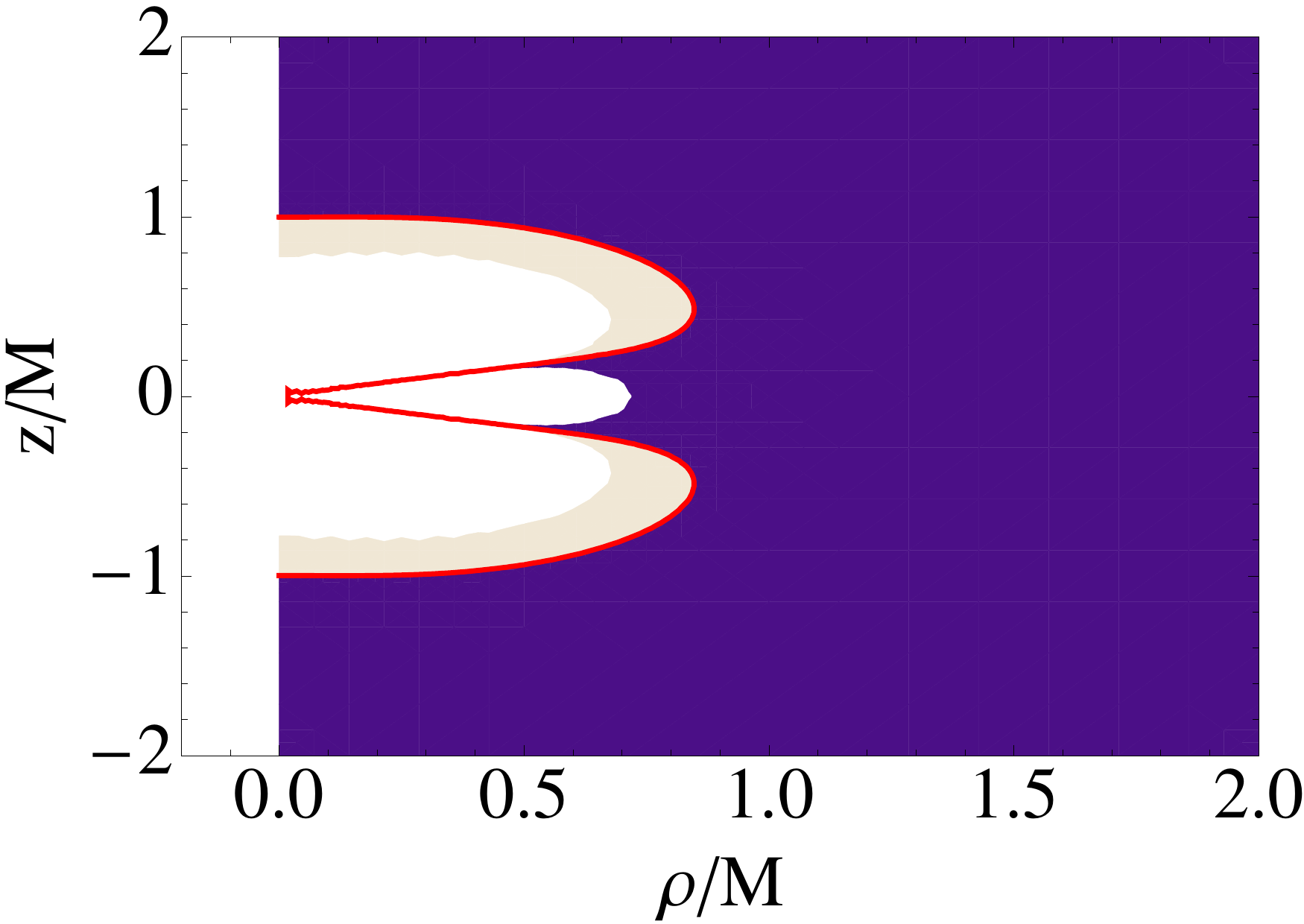}
\includegraphics[width=.25\textwidth]{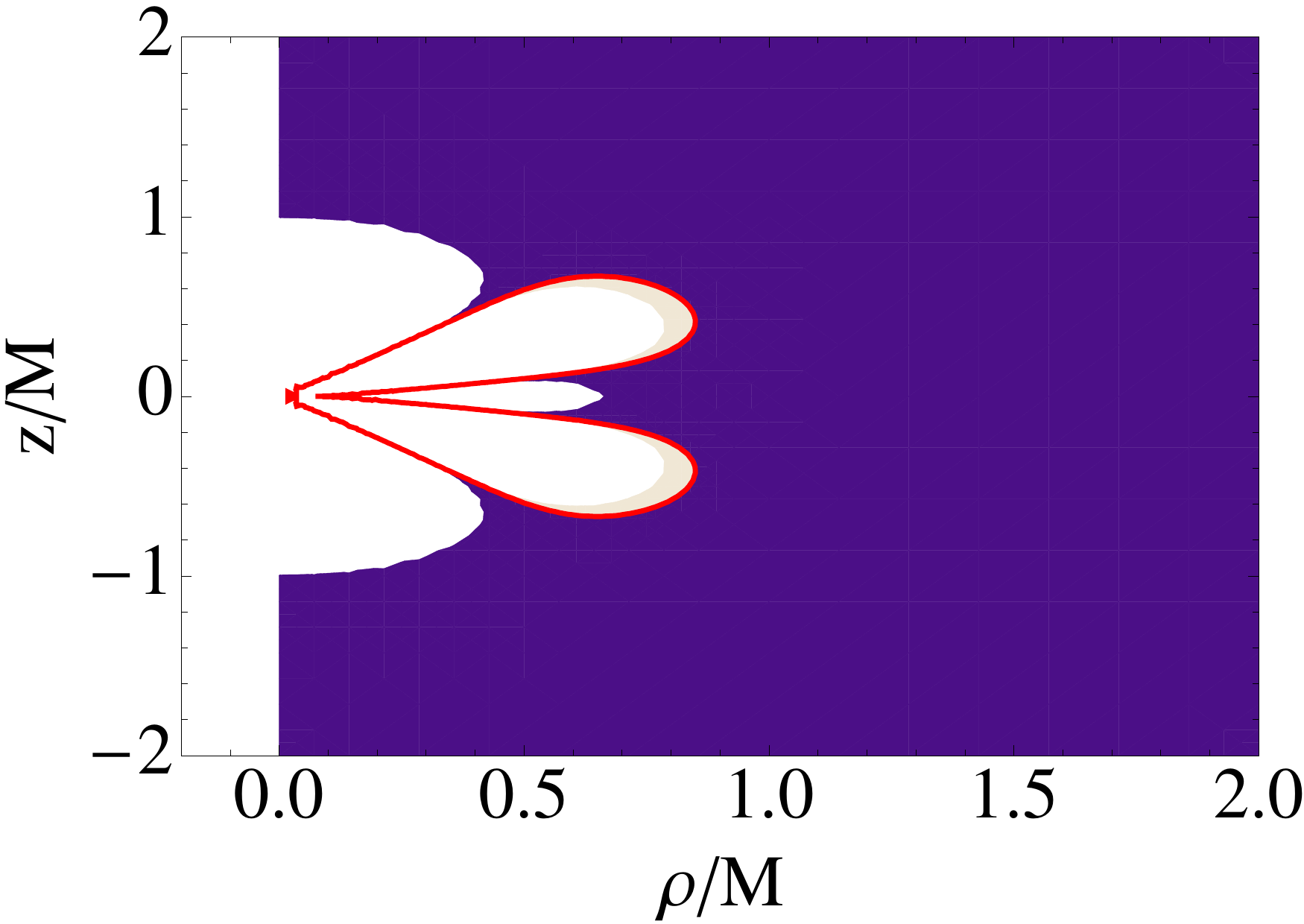} 
\includegraphics[width=.25\textwidth]{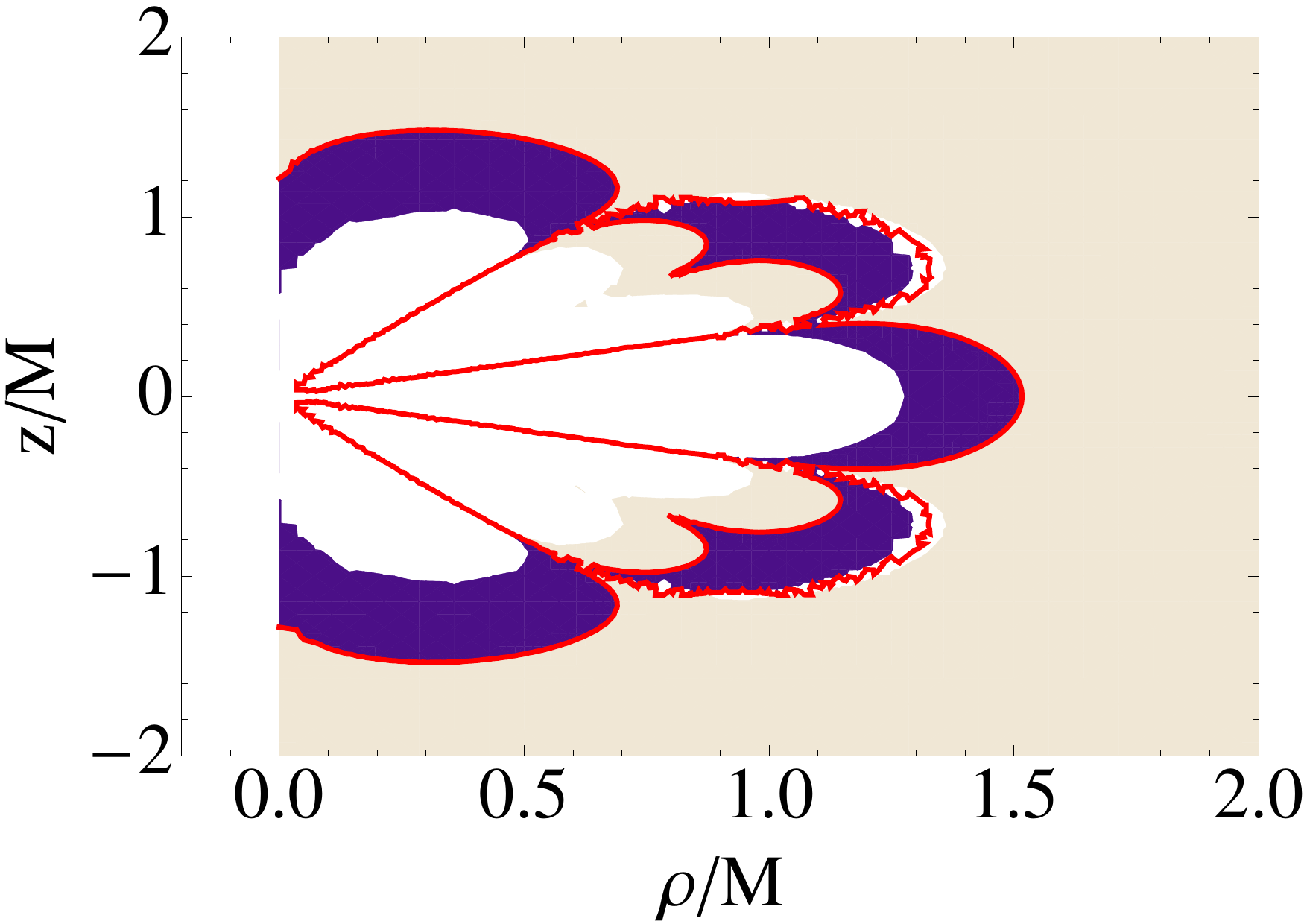} 
\includegraphics[width=.25\textwidth]{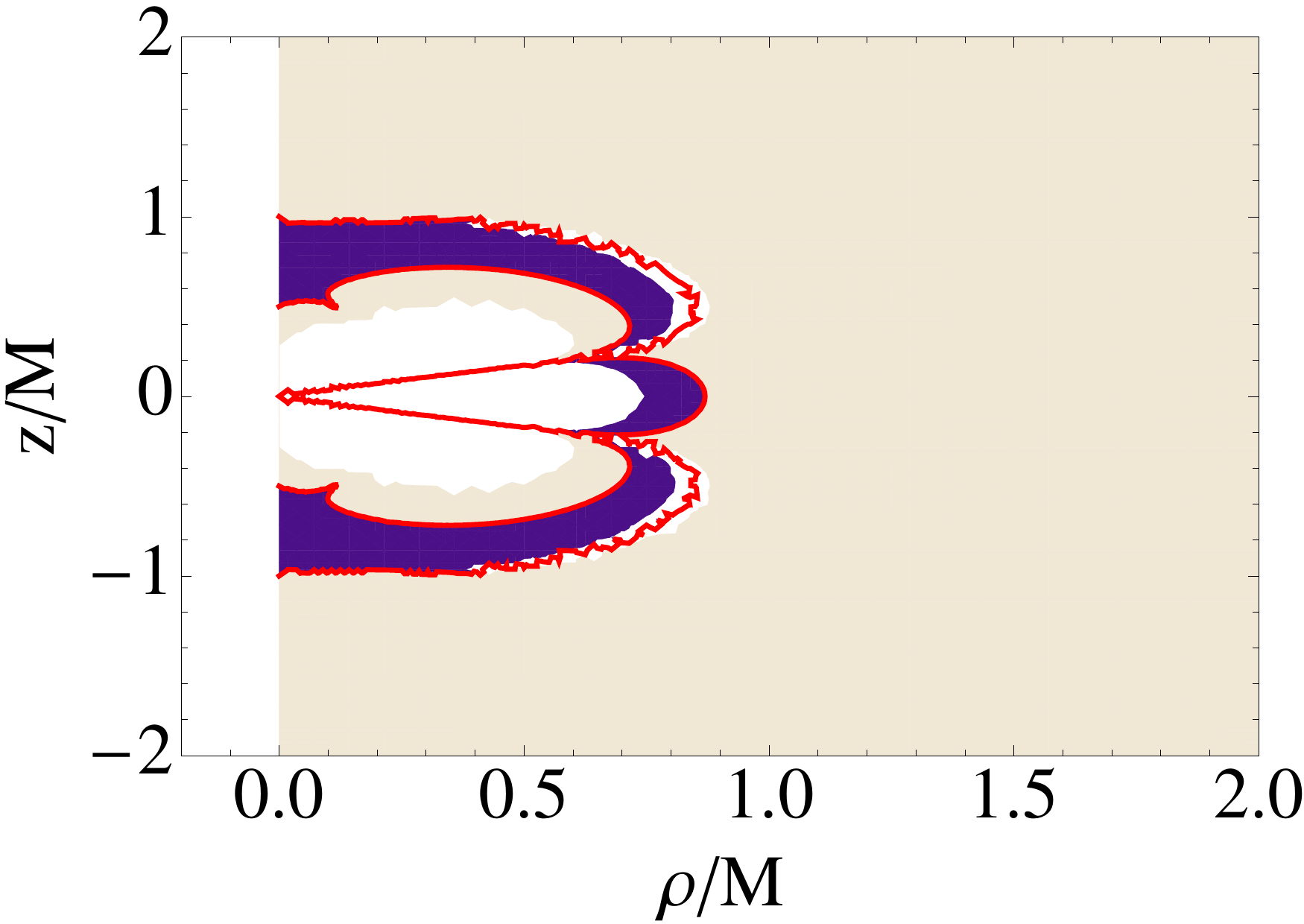}
\includegraphics[width=.25\textwidth]{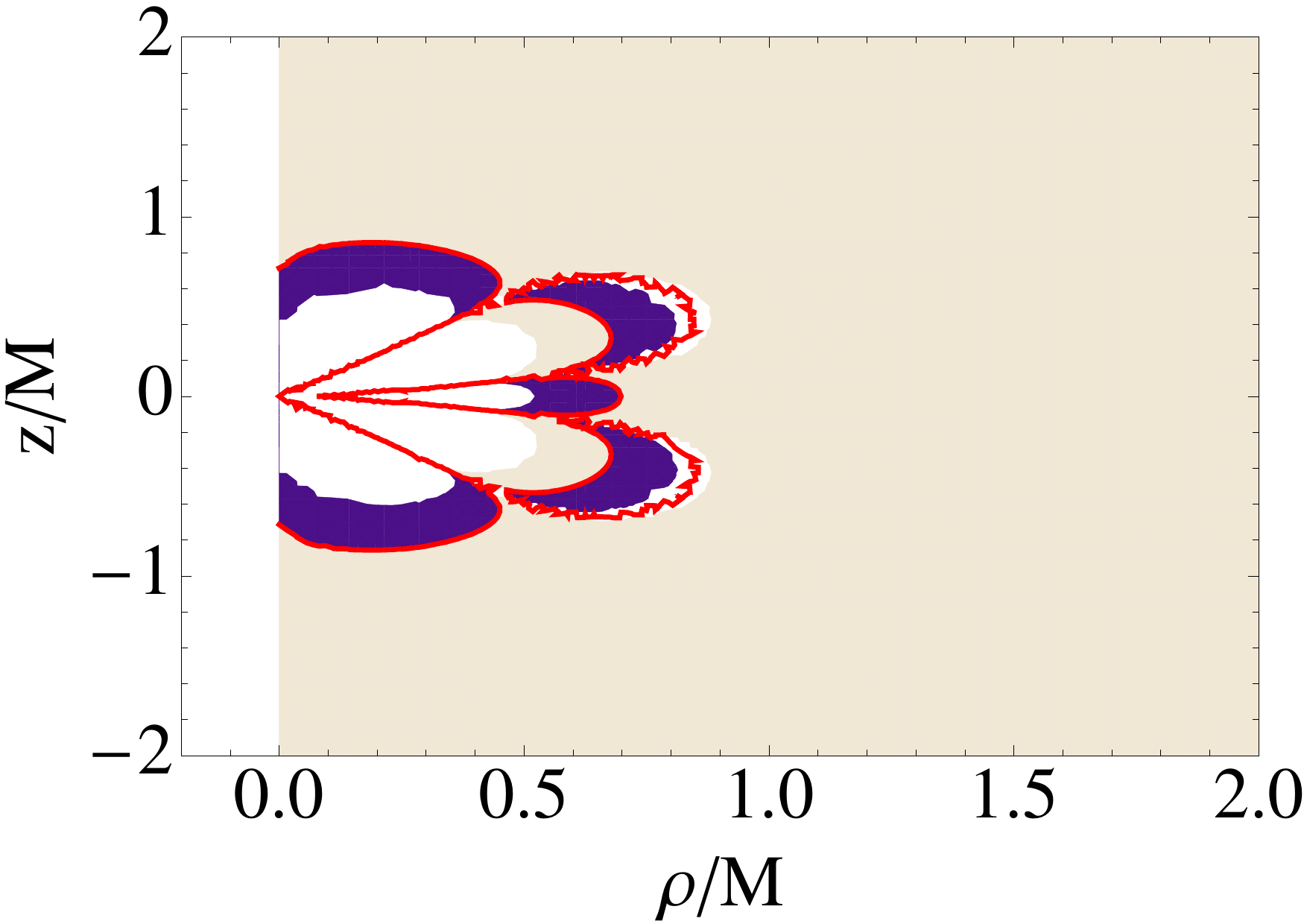} 
\includegraphics[width=.25\textwidth]{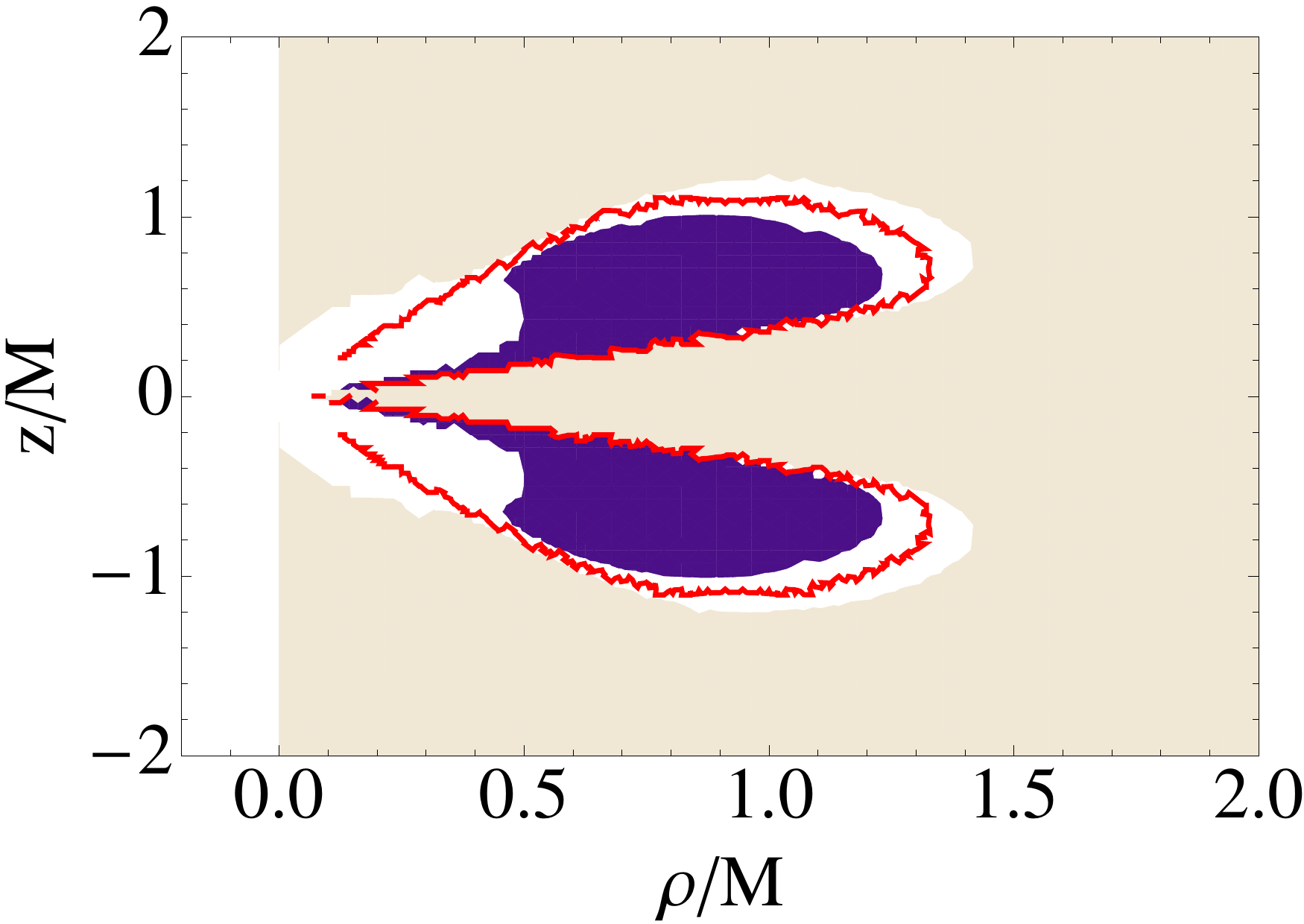} 
\includegraphics[width=.25\textwidth]{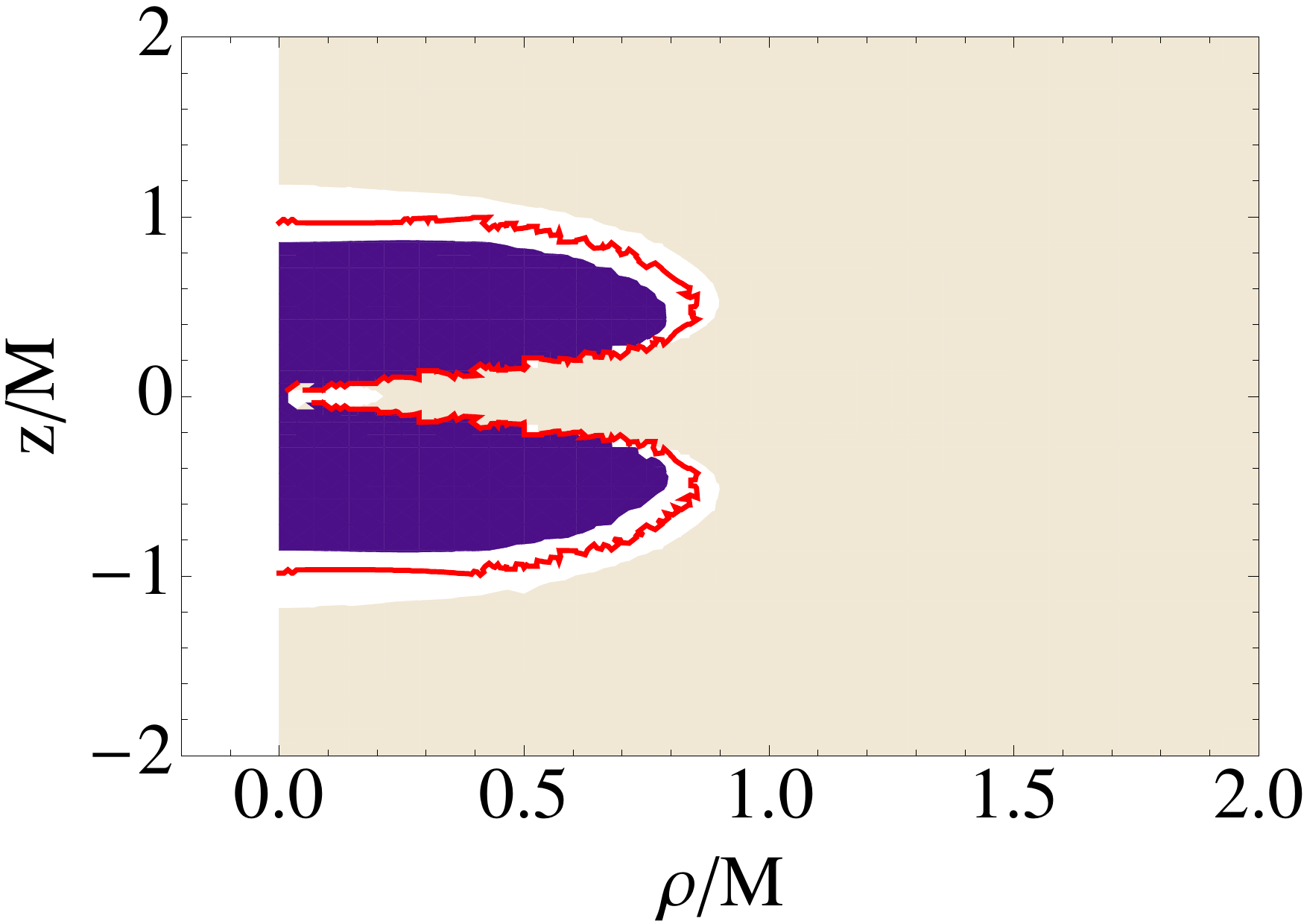}
\includegraphics[width=.25\textwidth]{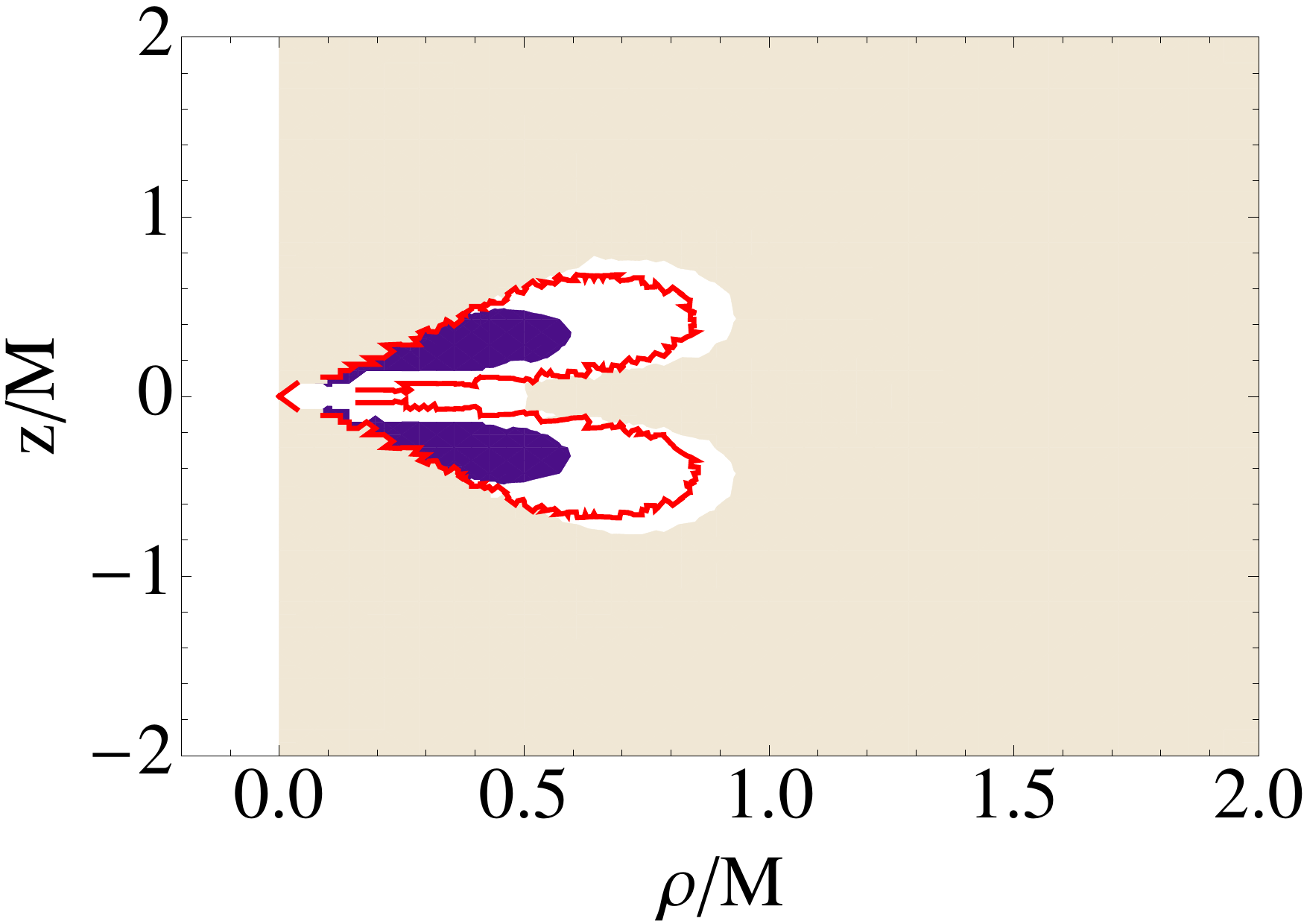} 
\caption{Typical plots of the behaviour of the metric functions $g_{tt}$ (top row), $g_{\varphi\varphi}$ (middle row), and $g_{\rho\rho}$ (bottom row), for different values of the spin parameter $j$ and the quadrupolar deformability $\alpha$ ($j=0.5,\, \alpha=8$ for the left column, $j=0.25,\, \alpha=5$ for the middle column, and $j=0.125,\, \alpha=3$ for the right column). The red lines indicate the locations where the metric functions are zero. The regions where the functions are negative are with dark colour, while the region where the functions are positive are with beige colour. Finally the blank regions indicate regions where the functions have singular behaviour. }
\protect\label{patholog}
\end{figure*}

With the spacetime at hand we should briefly investigate some of its properties, to make sure that there are no pathologies that would make the spacetime unsuitable for NSs. Such problematic properties would be the presence of singularities, horizons or closed timelike curves (CTCs). Furthermore one could also check for the existence of ergoregions.   

With respect to horizons, by construction in Weyl-Papapetrou coordinates, a horizon will be located at coordinate $\rho=0$. This is due to the following reason. A horizon is the boundary that separates the region of space where there can exist stationary observers from the region where there cannot. When the spacetime is stationary and axisymmetric, there exist two Killing vectors associated to the symmetries, one timelike and one spacelike. In this case, the fourvelocity for a general stationary observer can be written as a linear combination of the timelike and the spacelike Killing vectors, of the form
\be u^{\mu}=\lambda \left( \xi^{\mu}+\Omega \eta^{\mu}  \right), 
\ee     
where $\xi^{\mu}$ is the timelike and $\eta^{\mu}$ is the spacelike Killing vector, while $\Omega=d\varphi/dt$ is the observer's angular velocity and $\lambda$ is a normalisations factor so that $g_{\mu\nu}u^{\mu}u^{\nu}=-1$, i.e., the fourvelocity is timelike.\footnote{Essentially this $\lambda$ is the redshift factor, $dt/d\tau$, between coordinate time and proper time.} This leads to the constrain for $\lambda$,
\be \lambda^{-2}=-g_{tt}-2\Omega g_{t\varphi}-\Omega^2 g_{\varphi\varphi} >0. \ee
If $g_{\varphi\varphi}$ is positive (negative), the expression takes positive values only if $\Omega$ is between (outside) the two roots of the quadratic expression and the horizon is at the coordinates for which there is only one real root for $\Omega$, i.e., when the discriminant is zero, or $(g_{t\varphi})^2-g_{tt}g_{\varphi\varphi}=0$. In Weyl-Papapetrou coordinates, this is the definition of the coordinate $\rho^2$, therefore any horizon would correspond to $\rho=0$.

The boundary of the ergoregion, or the static limit, i.e., the point beyond which there cannot be any non-rotating observers is given by the condition of having $g_{tt}=0$, while the boundary for the region of CTCs is given by the condition $g_{\varphi\varphi}=0$. Since the $g_{\rho\rho}=g_{zz}$ components of the metric depend on $f^{-1}=(-g_{tt})^{-1}$, the condition for the ergoregion boundary could also be related to possible singularities. For this reason we will plot the curves on the $(\rho,z)$ plane which correspond to these conditions being satisfied. In Figure \ref{patholog} we have plotted the behaviour of the metric functions for indicative values of $j$ and $\alpha$. The metric functions are generally well behaved outside the region $|\rho/M,z/M|\lo2$ (the possible pathologies are well inside that region), which is always inside the surface of a NS for the various EoSs that we have investigated, therefore the given metric will be able to describe the exterior of NSs without problems. 
Further analysis of the spacetime properties and these pathologies is beyond the scope of this work.

\section{Comparison of the spacetime against NS spacetimes}
\label{sec:4}

\begin{figure*}
\centering
\includegraphics[width=.32\textwidth]{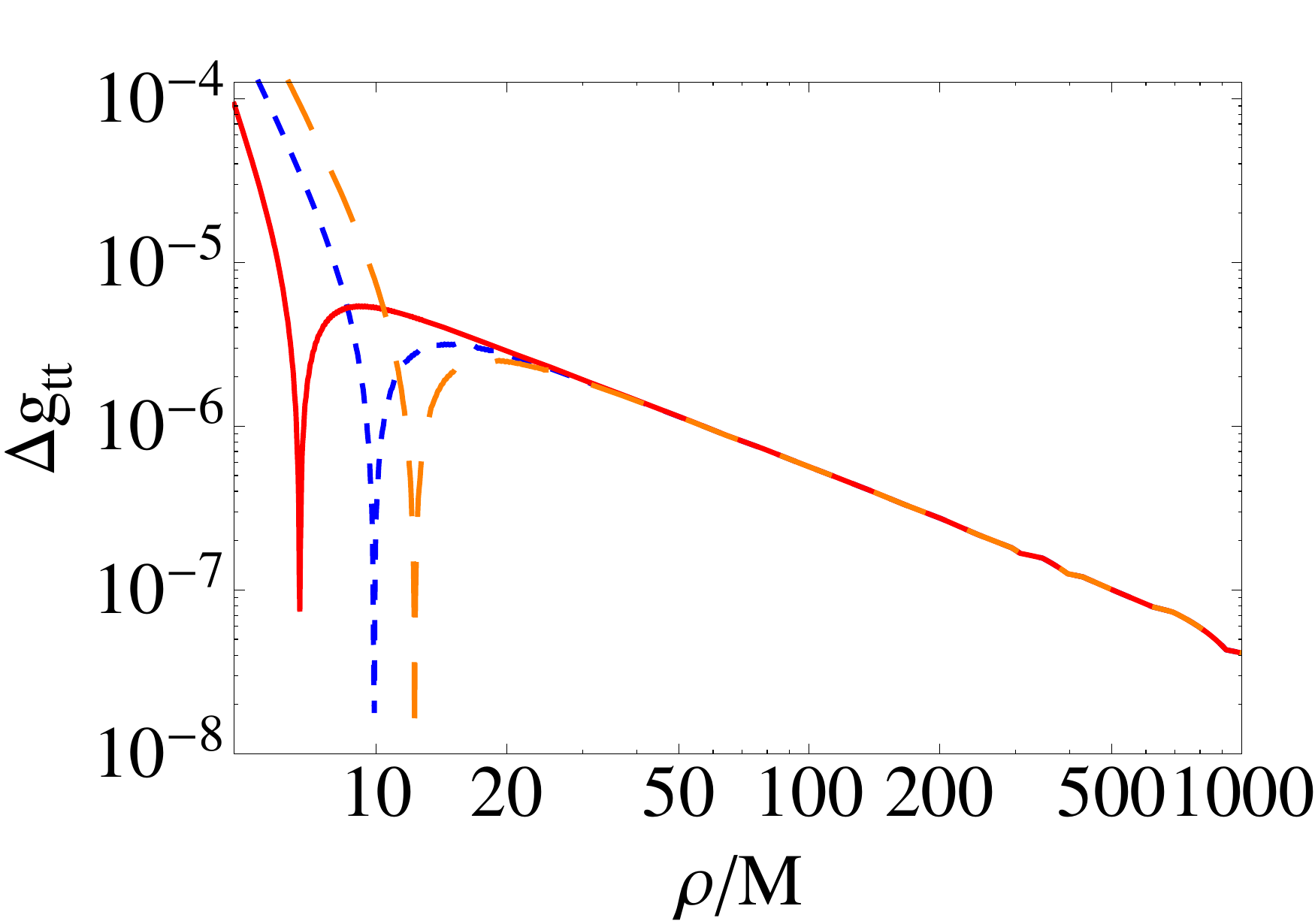} 
\includegraphics[width=.32\textwidth]{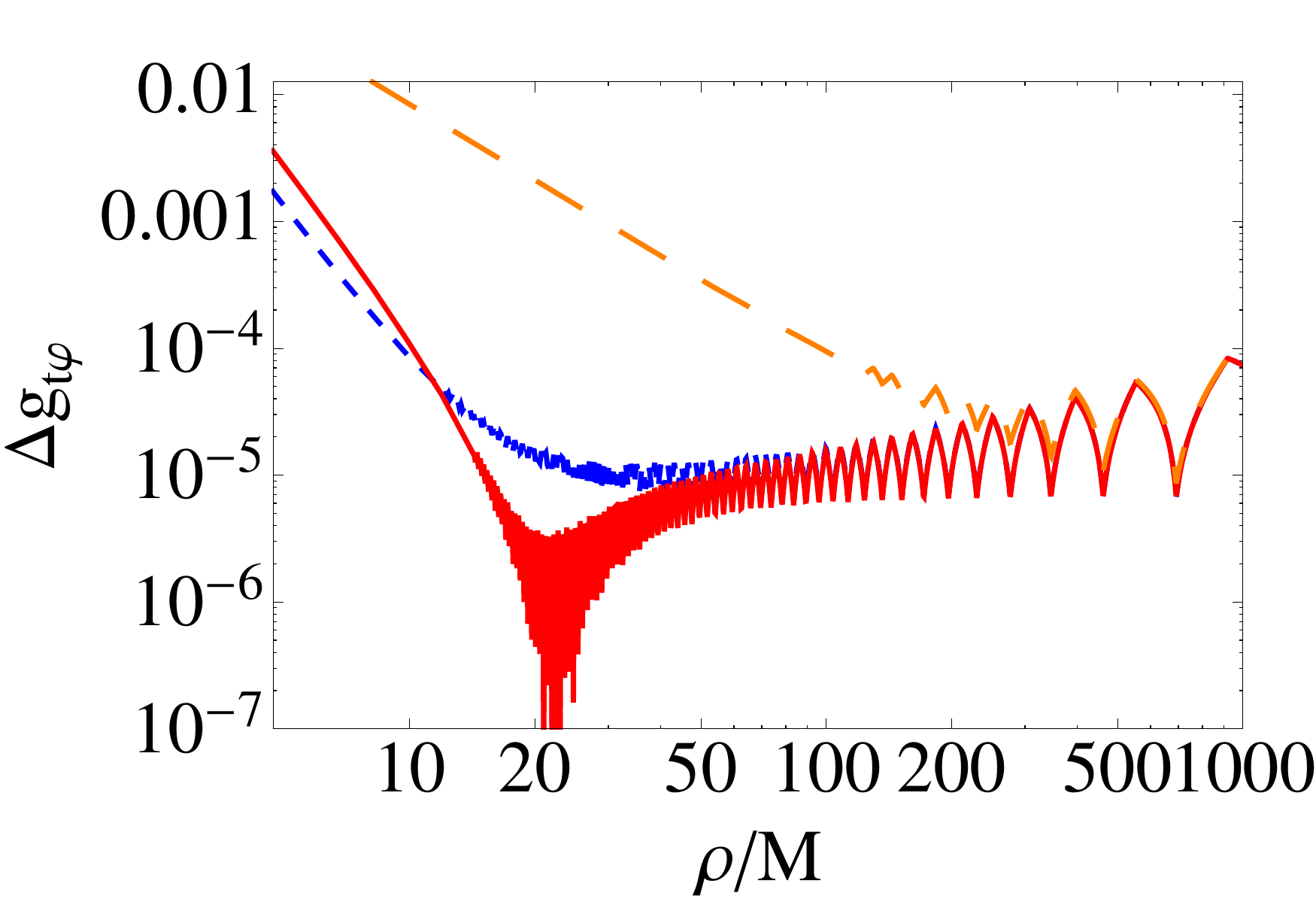}
\includegraphics[width=.32\textwidth]{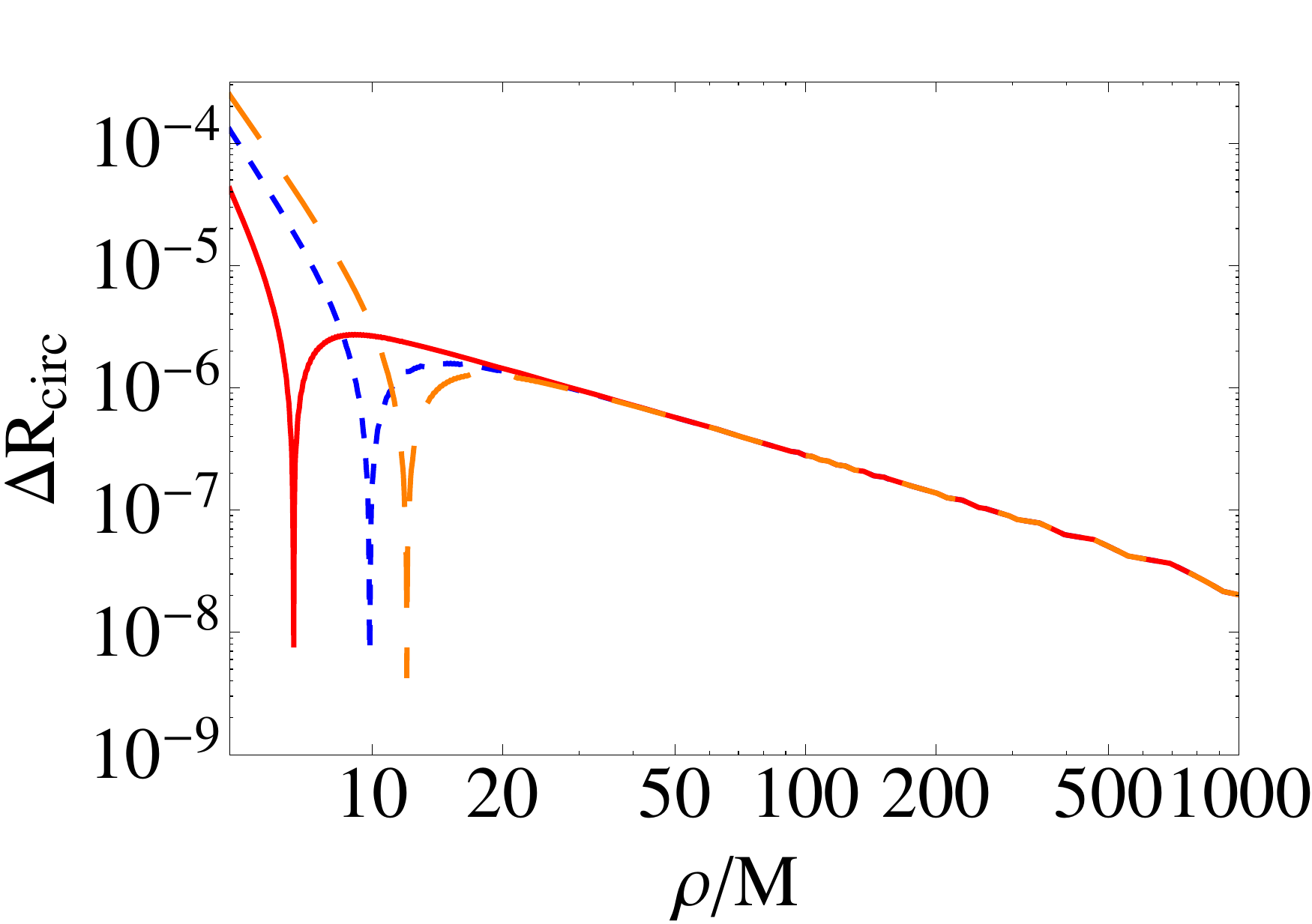} 
\includegraphics[width=.32\textwidth]{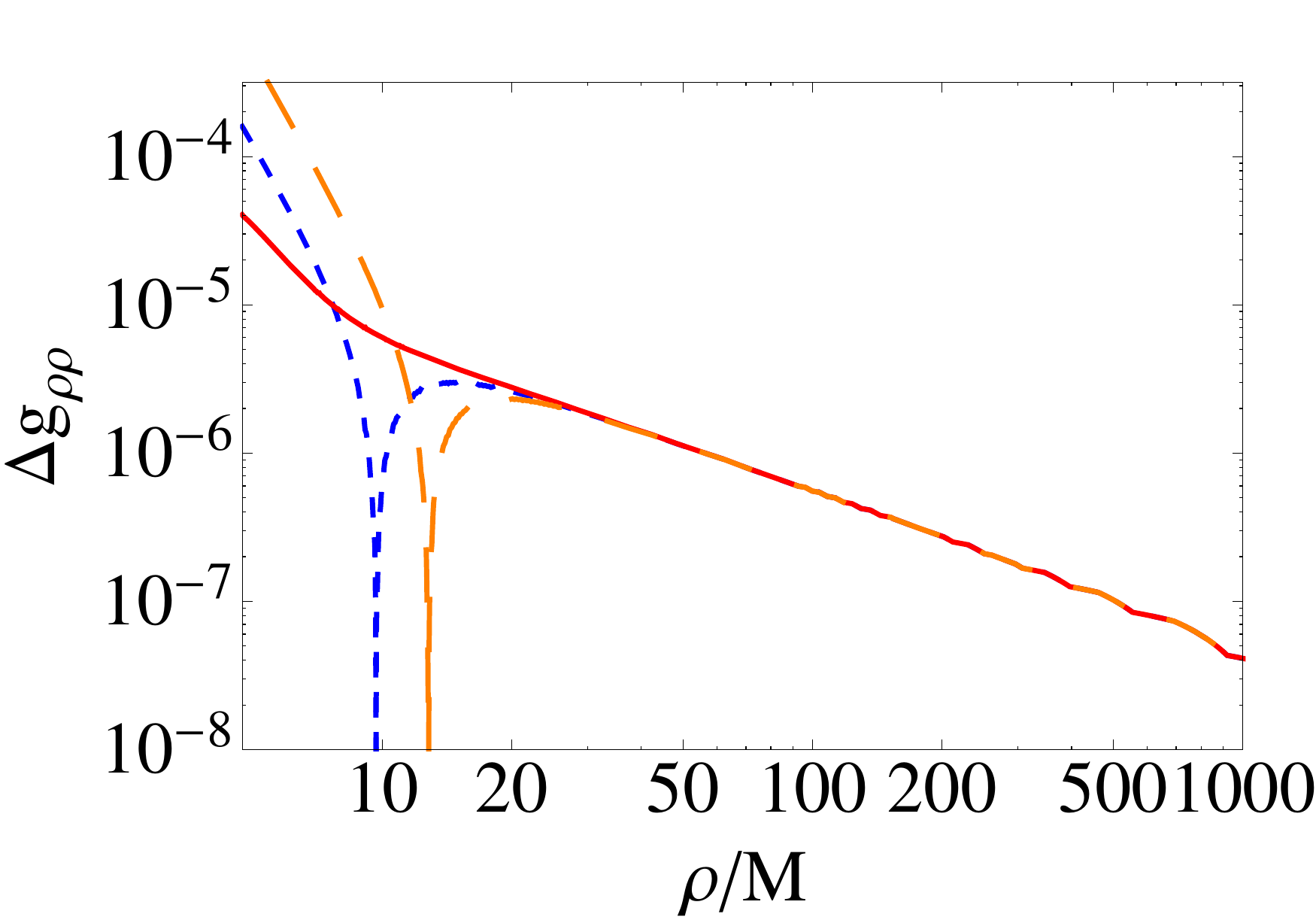} 
\includegraphics[width=.3\textwidth]{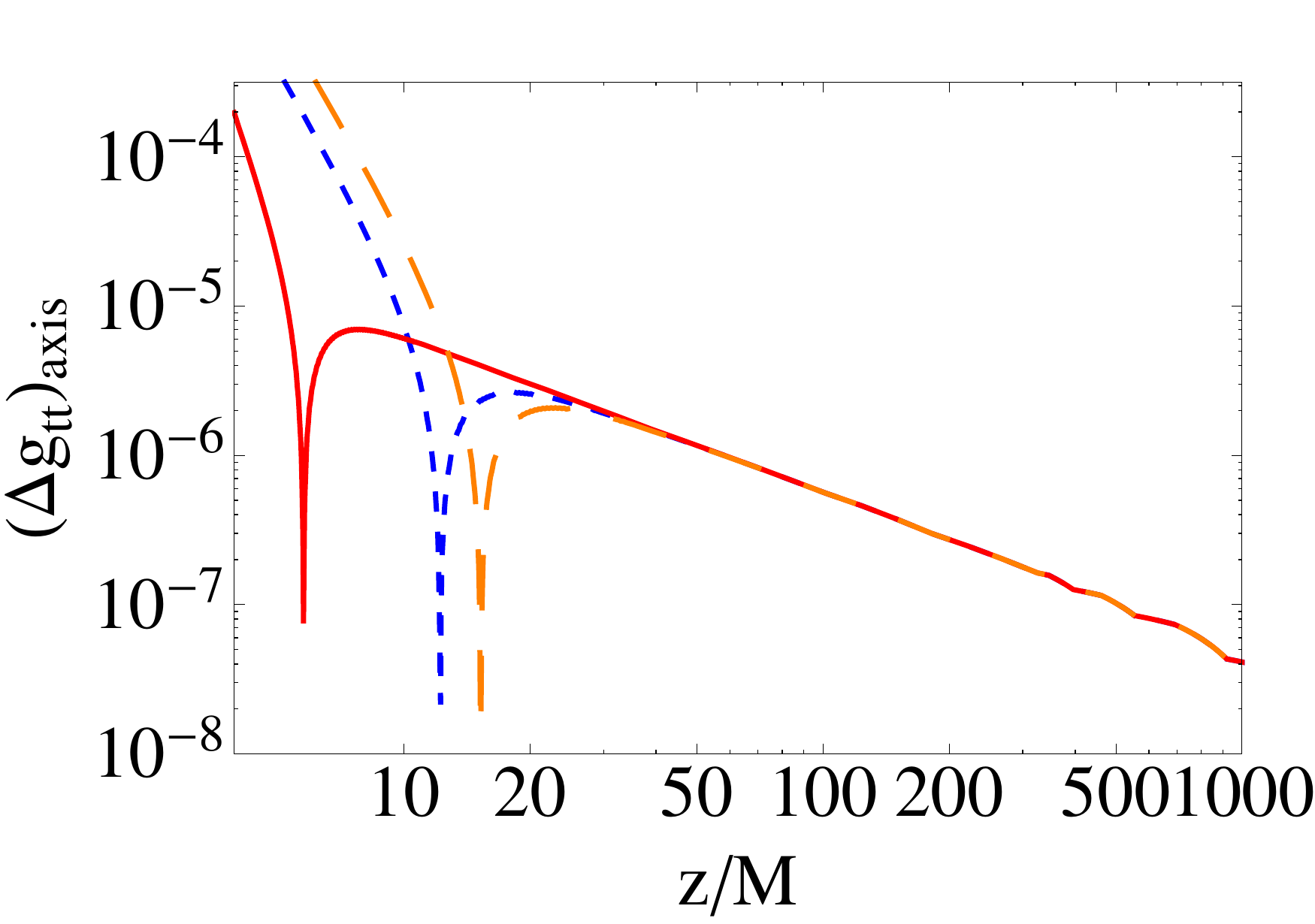}
\caption{Typical relative difference plots for the metric functions on the equatorial plane and along the axis of symmetry. The plots are made using EoS FPS for a numerical model of $M=1.4M_{\odot}=2.0876\textrm{km}$ rotating with a spin parameter of $j=0.453$ and having $\alpha= 4.209$. The plots show three curves which correspond to the metric proposed here (red solid curve), the two-soliton spacetime (blue dotted curve), and the Hartle-Thorne metric (orange dashed curve).
}
\protect\label{metricf}
\end{figure*}

We now proceed to test how good an approximation of the spacetime around rotating NSs the analytic spacetime presented here is. To do this we compare the analytic spacetime against numerical spacetimes produced with the RNS code \citep{Sterg} for various EoSs. The models and the EoSs that we are using are the same as the ones used by \cite{twosoliton} for the evaluation of the two-soliton spacetime \citep{twosoliton2}, which are the same as some of the models used by \cite{berti-stergioulas} for the evaluation of the \cite{Mankoetal} solution. This is done because these previous investigations form a baseline for the comparison which we will take advantage of here. For comparison purposes we will also use the \cite{HT} exterior solution. 

The NS models that we are using are constructed with three EoSs of varying stiffness. These are, the AU EoS (soft), the FPS EoS (moderate stiffness) and the L EoS (stiff). One can find details for these EoSs in the work of \cite{CST} or in the supplemental material of \cite{Pappas:2013naa} and the references therein. The models used are sampling the parameter space of $j$ and $\alpha$ covering the range from small values of $j$ up to maximum spin (at the Kepler limit) and high values of $\alpha\sim8$ (lower mass models) down to low values of $\alpha\sim1.5$ (corresponding to models close to and above the maximum non-rotating mass).          

The quantities that we are using for the comparison of the spacetimes are the metric functions themselves and quantities that have to do with the properties of the geodesics of the spacetime, which are the radius of the innermost stable circular orbit (ISCO) and the various characteristic frequencies, i.e., the orbital and the epicyclic or the precession frequencies \citep[see][for details]{twosoliton}. For the comparison, an analytic spacetime is identified to a numerical spacetime by assuming the same multipole moments (which in the case of the present work this is done using the mass, the spin, and the quadrupole that correspond to a given numerical model while for the spin octupole and mass hexadecapole the universal relations (\ref{uniRelS3},\ref{uniRelM4}) are used). 

Figure \ref{metricf} shows plots of the relative difference of the analytic metric function from the numerical metric function 
\be \Delta g_{ab}=\left|\frac{g^N_{ab}-g_{ab}}{g^N_{ab}} \right| \ee
as a function of $\rho/M$ on the equatorial plane or as a function of $z/M$ on the axis of symmetry. 

The spacetimes that are used for the comparison are the two-soliton spacetime (blue dotted), the Hartle-Thorne exterior spacetime (orange dashed), and the approximate spacetime proposed here (red solid). The comparison is done in Weyl-Papapetrou coordinates and the Hartle-Thorne and numerical spacetimes are transformed to these coordinates first \citep[see][for details on the transformation]{berti-stergioulas,Pappas2008CQG,twosoliton}. The plots in Figure \ref{metricf} show the comparison between the $g_{tt}$, $g_{t\varphi}$, $R_{circ}=\sqrt{g_{\varphi\varphi}}$, and $g_{\rho\rho}=g_{zz}$ metric functions on the equatorial plane and the $g_{tt}$ metric on the axis of symmetry. In general one can see that the metric proposed here is as good as the two-soliton and in some cases even better, with the relative difference being in the order of $10^{-2}-10^{-3}$ for $g_{t\varphi}$ and $10^{-3}-10^{-4}$ for the rest of the functions.\footnote{For models with masses lower than the maximum non-rotating mass of any particular EoS and not close to the maximum rotation rate, the accuracy is usually better than that by an order of magnitude.} For some models that are rapidly rotating and have very small $\alpha\lo1.5-2$, i.e., for models with masses close to the maximum mass sequences of their respective EoSs, we observe that the accuracy of the proposed spacetime deteriorates.

\begin{figure*}
\centering
\includegraphics[width=.3\textwidth]{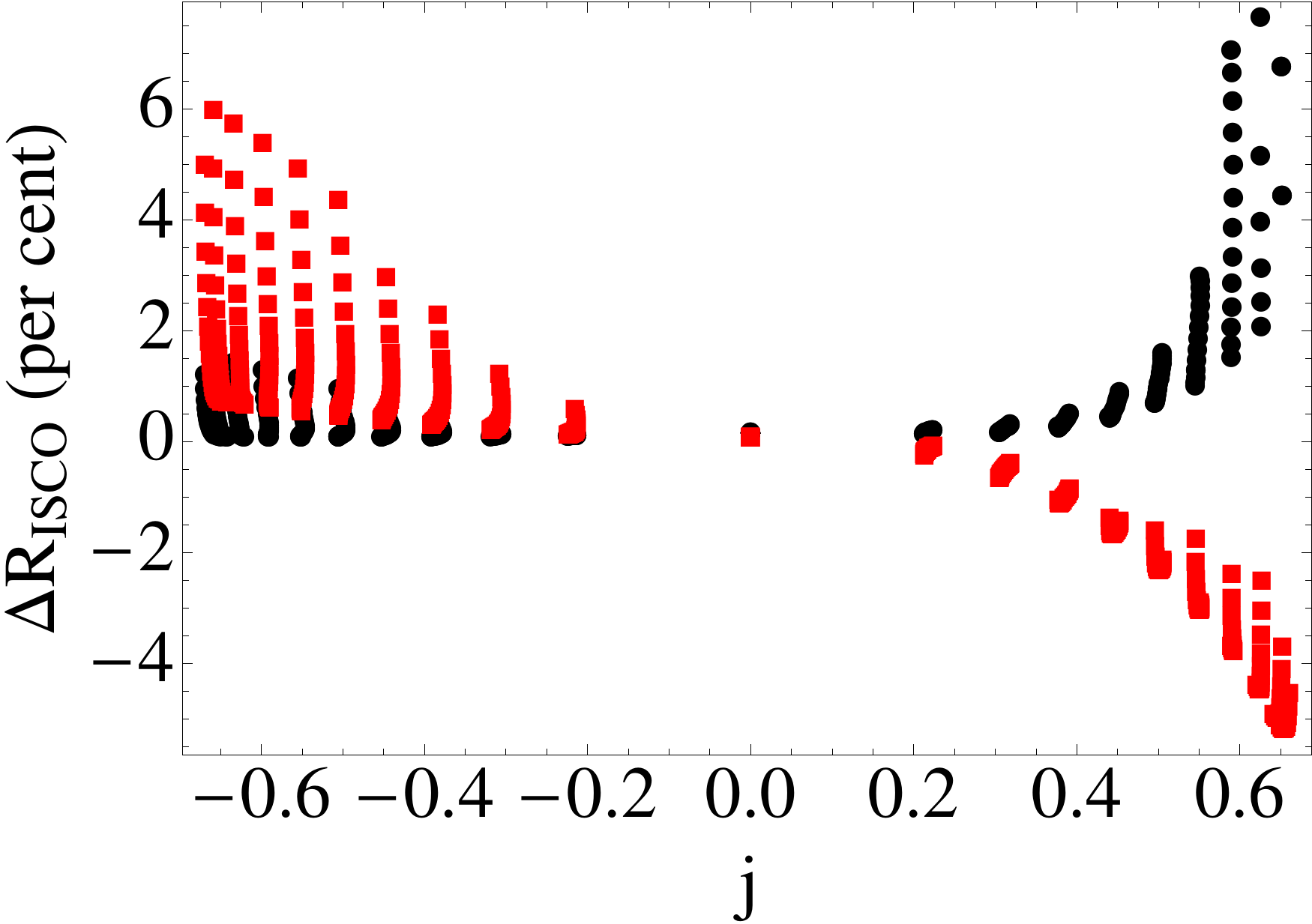} 
\includegraphics[width=.32\textwidth]{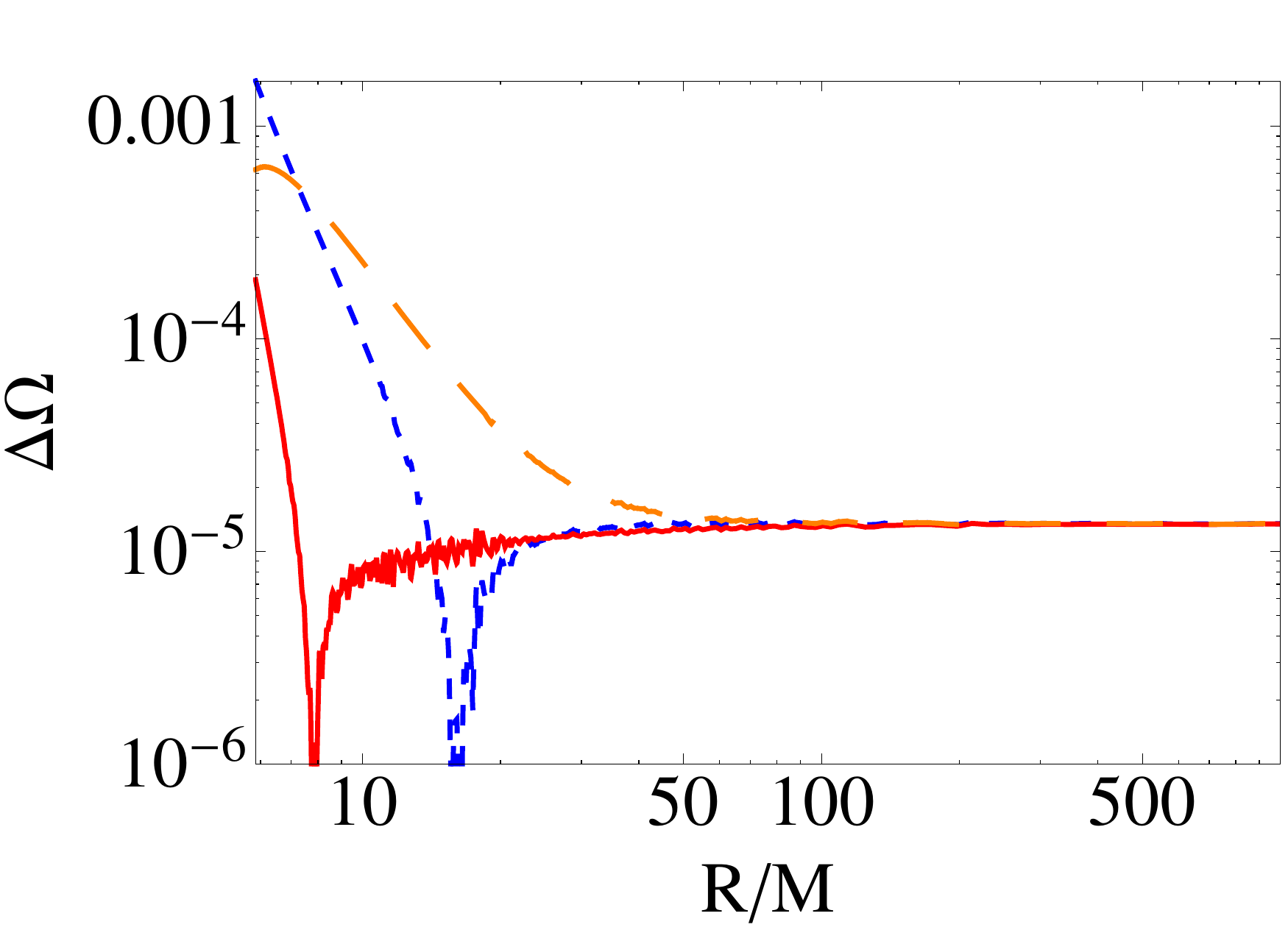}
\includegraphics[width=.32\textwidth]{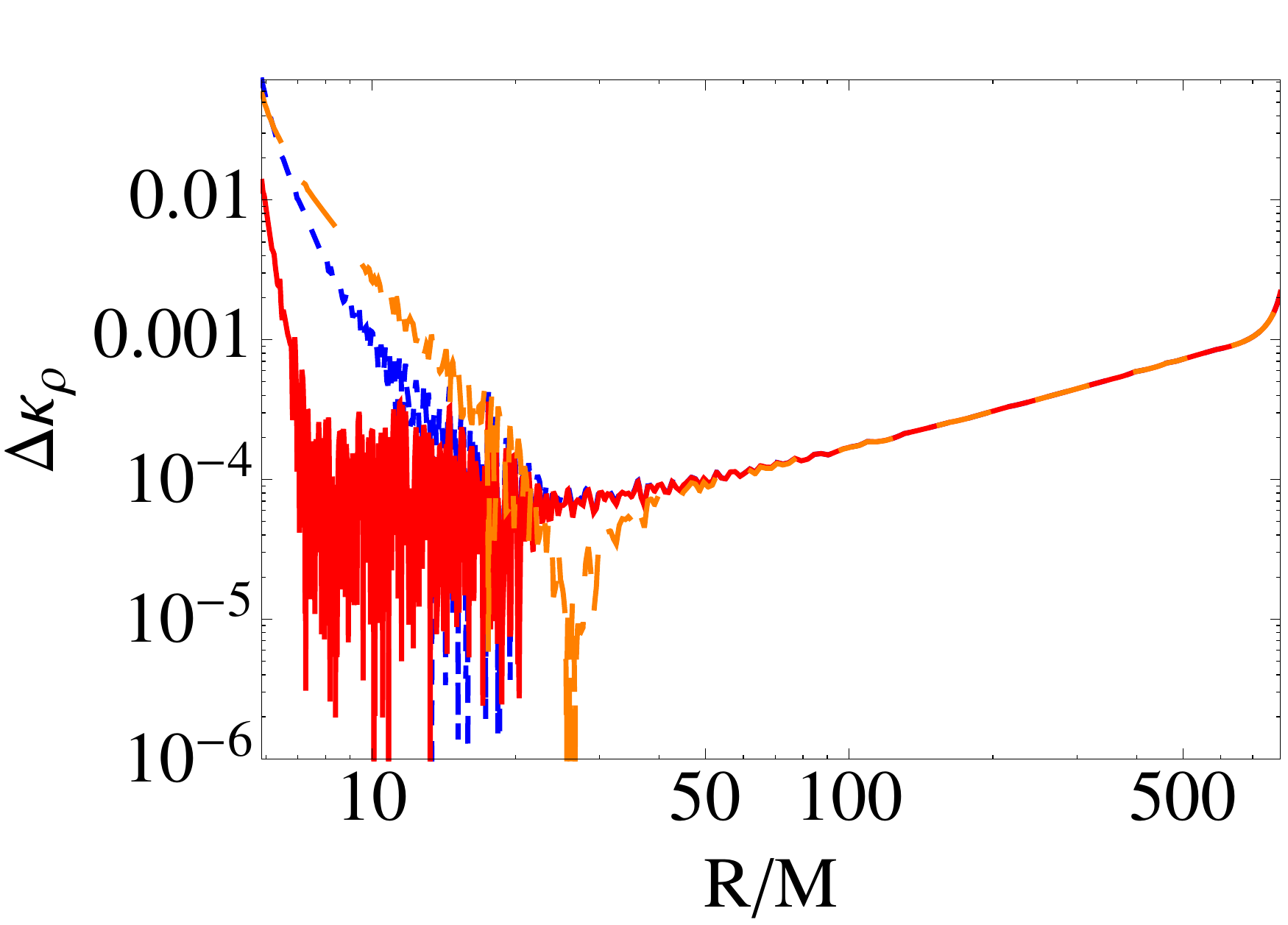} 
\includegraphics[width=.32\textwidth]{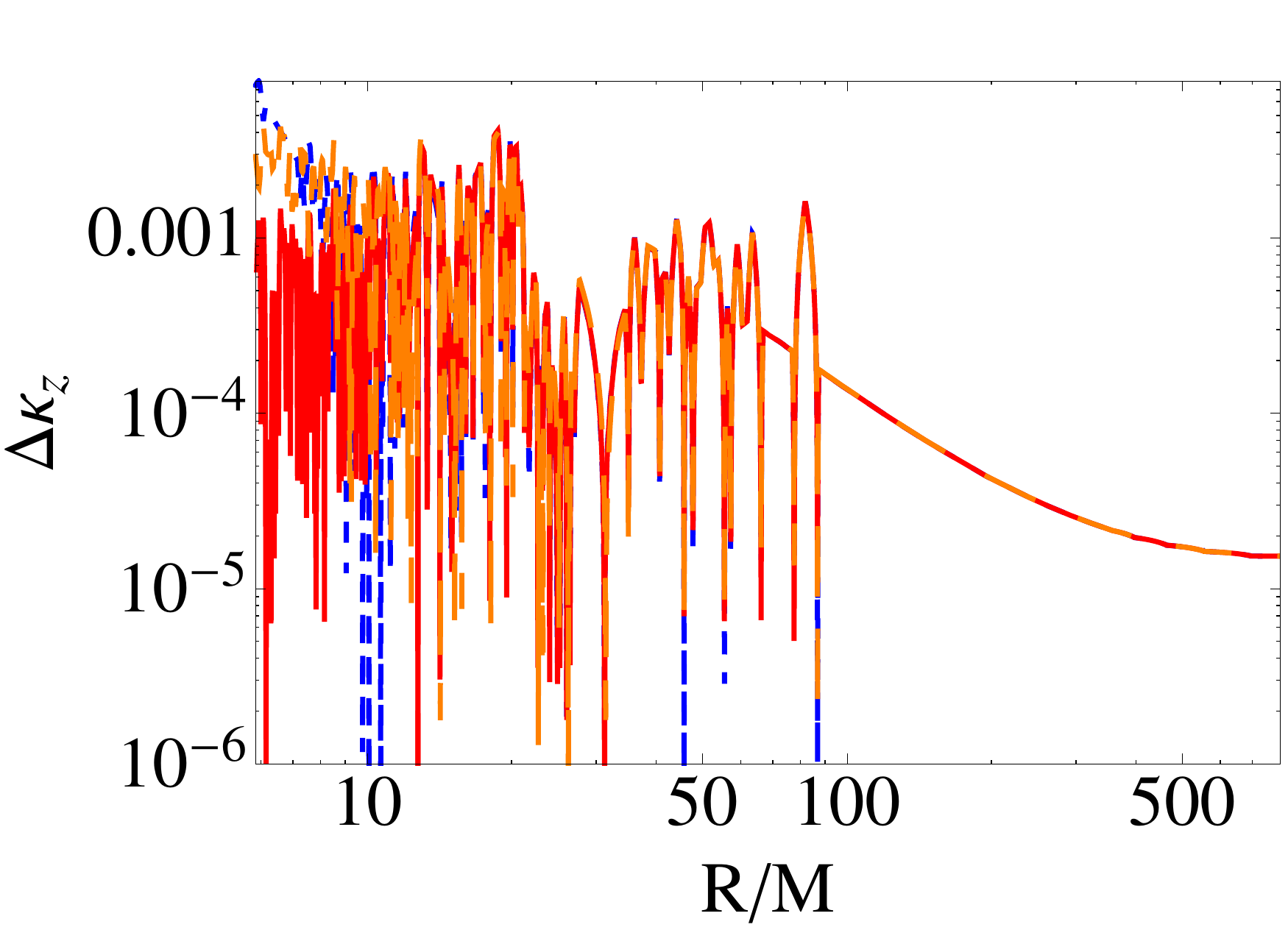} 
\includegraphics[width=.3\textwidth]{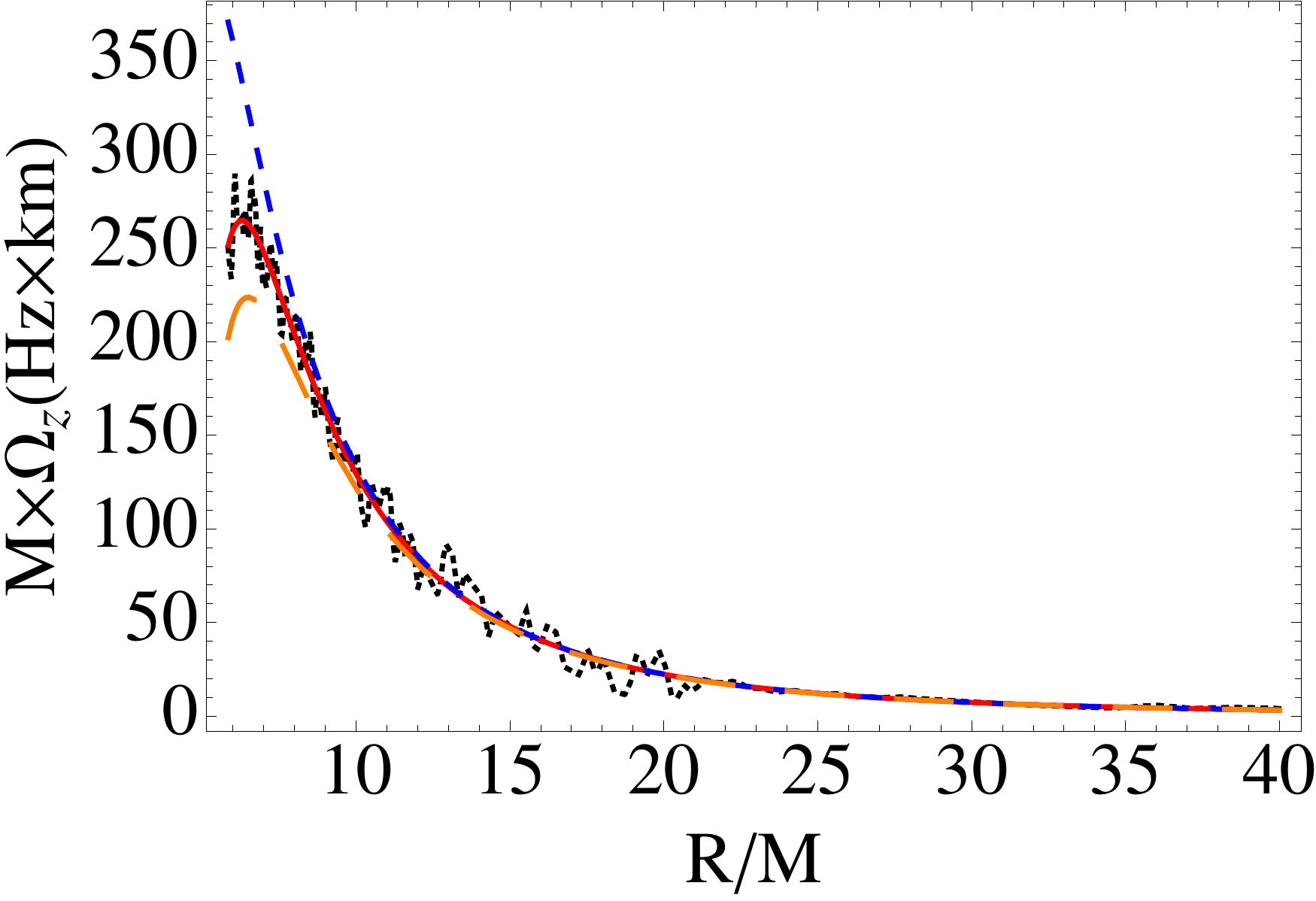}
\includegraphics[width=.32\textwidth]{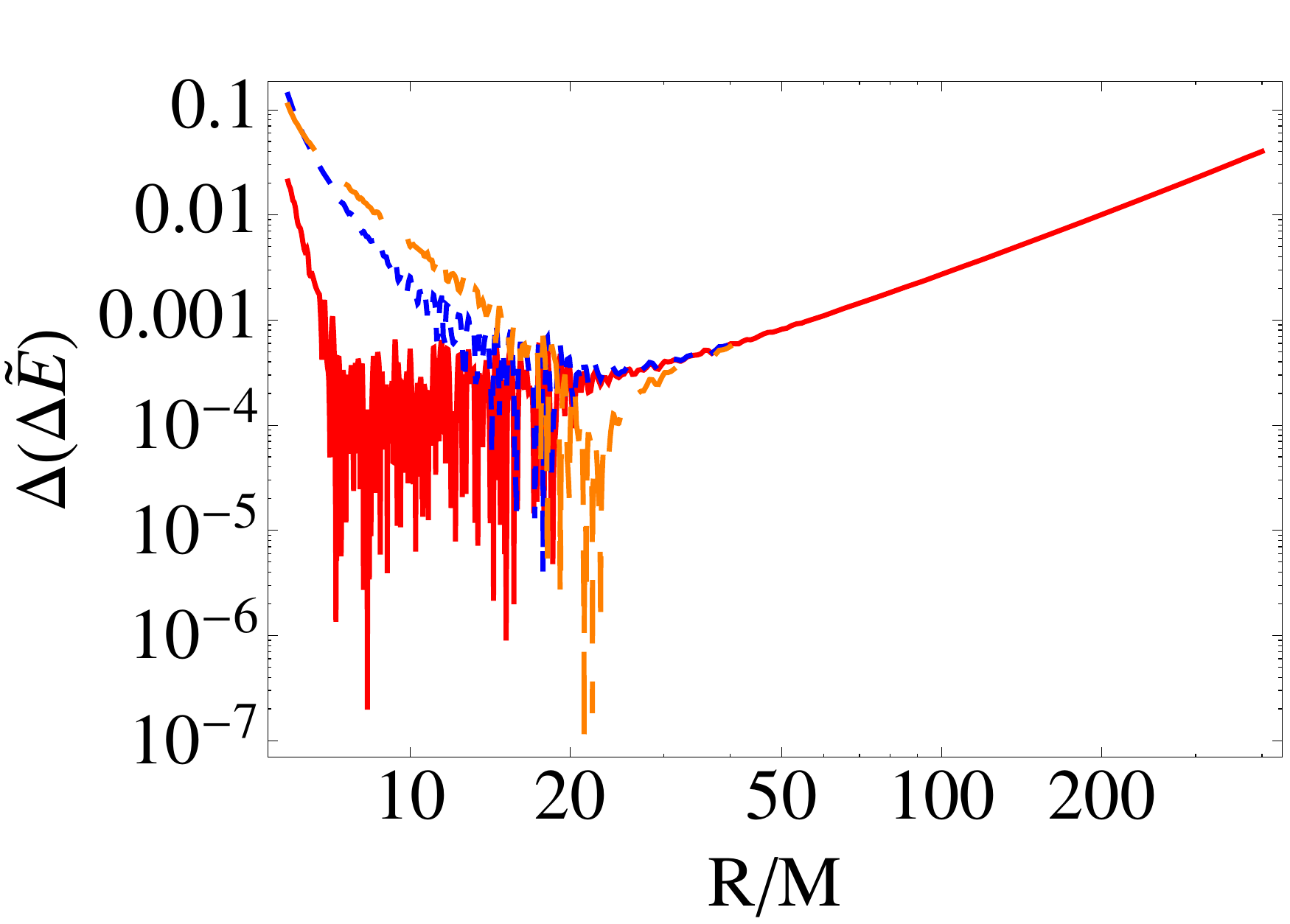}
\caption{Typical relative difference plots for the various geodesic properties of the spacetime between the numerical spacetime and the corresponding analytic spacetime. The top left plot gives the relative difference of the ISCO for the approximate metric (black circles) and the Hartle-Thorne metric (red squares). The models are constructed with the FPS EoS and we have plotted all the NS models that have an ISCO outside the surface of the star and for which the proposed metric has an ISCO (see discussion in the main text). The top middle plot shows the relative difference in the orbital frequency of circular equatorial orbits, $\Delta\Omega$, as a function of the circumferential radius over the mass, between the three analytic metrics and the numerical metric. The top right plot shows the relative difference for the radial oscillation frequency of radially perturbed orbits and the bottom left plot shows the relative difference for the vertical oscillation frequency of slightly off-equatorial orbits. The bottom middle plot shows the nodal precession frequency $M\times\Omega_z$ for the numerical and the analytic spacetimes (we remind that $\Omega_z=2\pi\nu_z$). Finally the bottom right plot shows the relative difference of $\Delta\tilde{E}$ between the numerical and the analytic spacetimes. The frequency and $\Delta\tilde{E}$ plots are constructed using the same model as in figure \ref{metricf}, but the results are similar for all the EoSs. The  curves correspond to the metric proposed here (red solid curve), the two-soliton spacetime (blue dotted curve), and the Hartle-Thorne metric (orange dashed curve). The nodal precession frequency plot (bottom middle) shows also the numerical frequency (black) which follows the proposed metric curve. 
}
\protect\label{frequencies}
\end{figure*}

With respect to the properties of the geodesics of the spacetime, we focus on circular equatorial orbits and their perturbations. Equatorial particle orbits on a Papapetrou spacetime follow the equation of motion 
\be -g_{\rho\rho} \left(\frac{d\rho}{d\tau}\right)^2=\left(1-\frac{\tilde{E}^2 g_{\varphi\varphi}+2\tilde{E}\tilde{L}g_{t\varphi}+\tilde{L}^2 g_{tt}}{\rho^2}\right)_{eq}\equiv V(\rho), \ee
where $\tilde{E}$ is the conserved energy per unit mass of the particle and $\tilde{L}$ is the conserved angular momentum with respect to the axis of symmetry per unit mass and everything is calculated on the equatorial plane for $z=0$. Circular orbits in this case satisfy the conditions $V(\rho)=0$ and $dV(\rho)/d\rho=0$, which are conditions for a minimum of the effective potential. The additional condition that the second derivative of the potential is also zero, i.e., $d^2V(\rho)/d\rho^2=0$, which correspond to a turning point, specify the location of the ISCO. Additionally, for circular geodesics the orbital frequency is given by the expression 
\be
\Omega(\rho)=\frac{-g_{t\varphi,\rho}+\sqrt{(g_{t\varphi,\rho})^2-g_{tt,\rho}g_{\varphi\varphi,\rho}}}{g_{\varphi\varphi,\rho}}, 
\ee
where the commas indicate the derivative with respect to the coordinate, while the energy and angular momentum per unit mass for these orbits is given in terms of the metric functions and $\Omega$ from the expressions,
\begin{align} \tilde{E}&=\frac{-g_{tt}-g_{t\varphi}\Omega}{\sqrt{-g_{tt}-2g_{t\varphi}\Omega -g_{\varphi\varphi}\Omega^2}},\label{energy}\\
         \tilde{L}&=\frac{g_{t\varphi}+g_{\varphi\varphi}\Omega}{\sqrt{-g_{tt}-2g_{t\varphi}\Omega -g_{\varphi\varphi}\Omega^2}}. \end{align}
From the energy per unit mass one can also define the energy change per logarithmic change in the orbital frequency as one goes from one circular orbit to the next,
\be 
\Delta\tilde{E}=-\Omega\frac{d\tilde{E}}{d\Omega}, \ee
which also characterises the circular equatorial orbits of a spacetime.

For general orbits, i.e., orbits that can be outside the equatorial plane, the previous equation of motion becomes
\be -g_{\rho\rho}\left(\frac{d\rho}{d\tau}\right)^2 -g_{zz}\left(\frac{dz}{d\tau}\right)^2=V(\rho,z),\label{eqmotion}\ee
and if we assume small perturbations around circular equatorial orbits along the radial direction or in the vertical direction, then we find that the perturbations have a harmonic behaviour with a radial and a vertical frequency which are given respectively as,
\begin{align} \kappa_{\rho}^2&=\left(\frac{d\tau}{dt}\right)^2 \left.\frac{g^{\rho\rho}}{2}\frac{\p^2 V}{\p\rho^2}\right|_c\,,\\
                                 \kappa_z^2&=\left(\frac{d\tau}{dt}\right)^2\left.\frac{g^{zz}}{2}\frac{\p^2 V}{\p z^2}\right|_c\,,\end{align}
as they are measured by an observer at infinity. The relevant precession frequencies will then be given by the difference $\Omega_a=\Omega-\kappa_a$, where $a$ is either $\rho$ for the periastron precession or $z$ for the nodal precession. 

We will use here all these properties of the geodesics to compare the approximate analytic spacetime against the numerical spacetime, since they constitute a test of how well the proposed spacetime captures the physical properties of NS spacetimes. 

The first property is the radius of the ISCO. The top left plot of figure \ref{frequencies} shows the relative difference of the numerical ISCO and the analytic ISCO of the proposed metric (black circles) and of the Hartle-Thorne metric (red squares) for models constructed with the FPS EoS for various rotations and masses. 
%
\begin{figure}
\centering
\includegraphics[width=.23\textwidth]{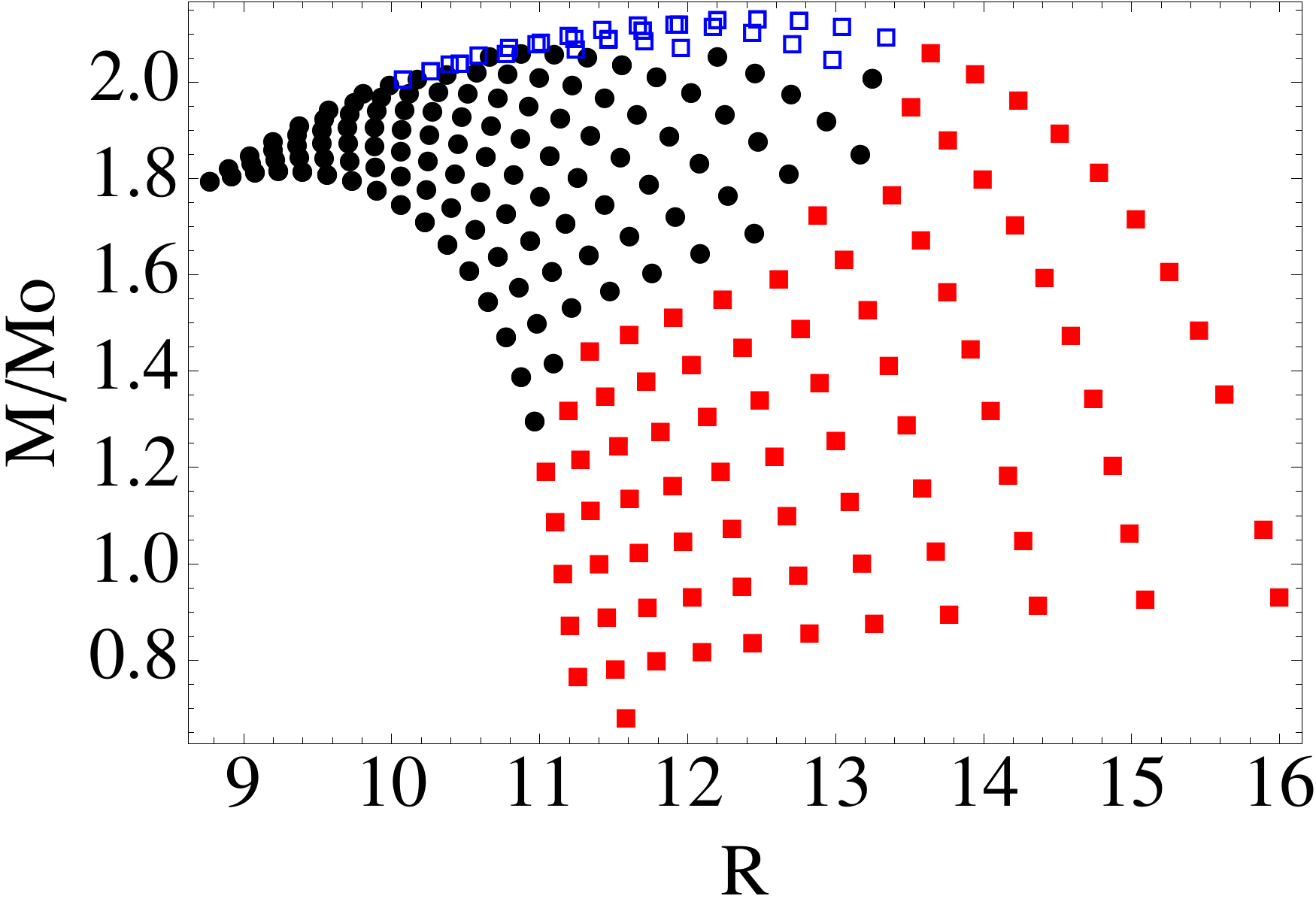}
\includegraphics[width=.23\textwidth]{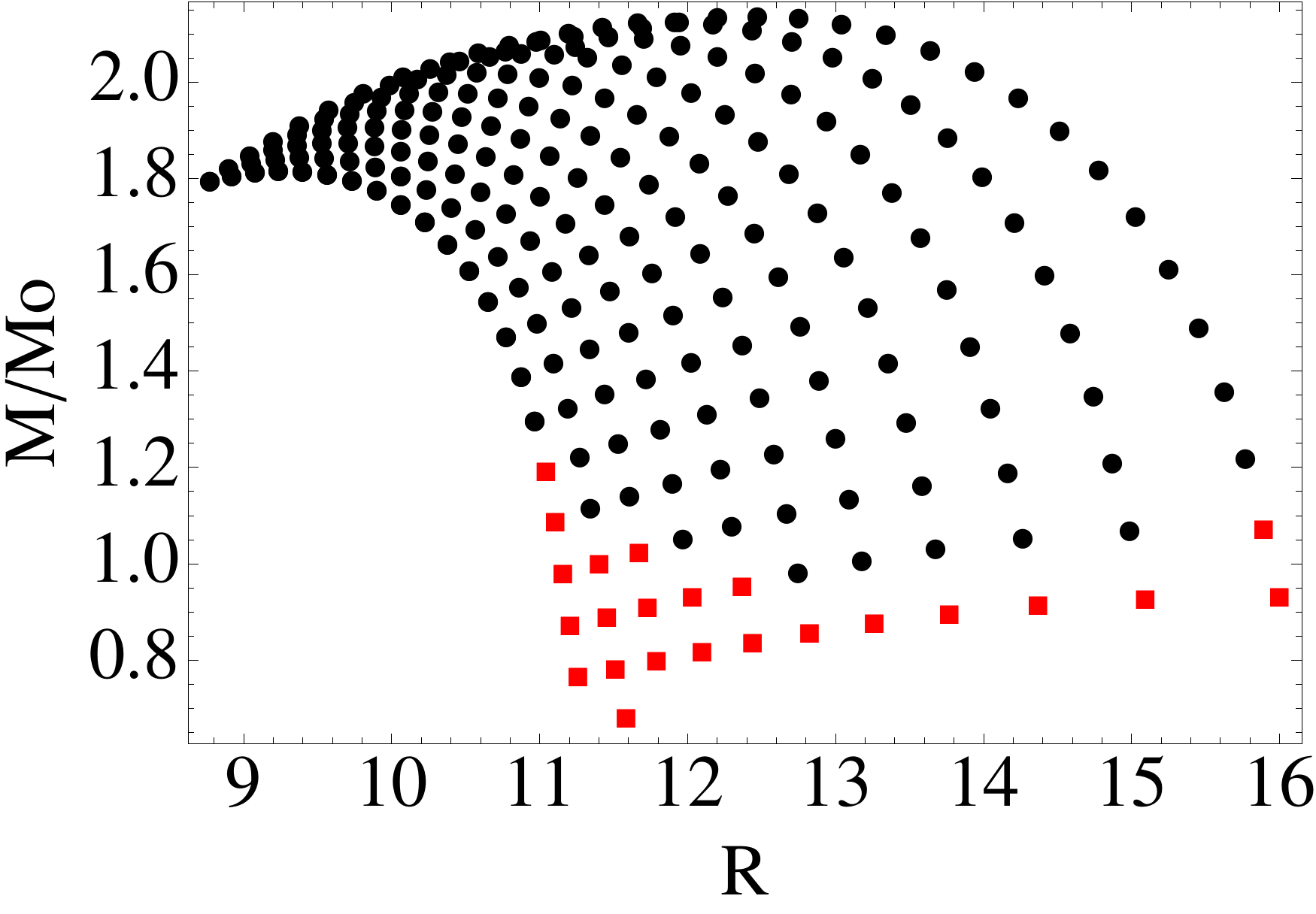}
\caption{Mass radius plot for the models used to calculate the ISCO in Fig. \ref{frequencies}, for the co-rotating case (left plot) and the counter-rotating case (right plot). The red squares correspond to models that have a NS radius larger than the ISCO, while the black circles correspond to models with radius smaller than the ISCO. The blue empty squares correspond to models for which the analytic spacetime does not have an ISCO or the relative difference is larger than 6 per cent.}
\protect\label{isco}
\end{figure}
%
One can see that the analytic metric gives an accurate ISCO radius for almost all of the models that an ISCO exists outside the surface of the star. For rapidly rotating models ($j>0.6$) and very low values of $\alpha$ ($\alpha\lo1.5$) the analytic metric has no co-rotating ISCO. This happens for the models close and beyond the maximum stable mass for some of the EoSs as figure \ref{isco} shows. This is not particularly problematic, since NSs that are rotating this rapidly and are this close to the maximum mass limit are not very likely to be observed.\footnote{\cite{Martinon2014PRD} have shown that cooling of proto-neutron stars results in zero temperature configurations with mass lower than the maximum allowed by an EoS, so the range of interest is between masses lower than that and down to the lowest well-constrained NS mass currently measured at $\sim1.0 M_\odot$-$1.1 M_\odot$ \citep{Lattimer2012, Martinez2015} - this, of course, ignores evolutionary considerations, like accretion of mass, but this may raise the mass by only up to $\sim 0.1 M_\odot$ \citep{Alpar1982, Tauris2012}.} Therefore we can say that it is safe to use this metric in general for $\alpha\go1.5-2$, which corresponds to models close to the maximum mass and up to $\alpha\lo10$, which corresponds for some EoSs to NSs with masses around $1.1M_{\odot}$.  

The next properties to compare are the various frequencies, i.e., the orbital frequency, the radial perturbation oscillation frequency, and the vertical perturbation oscillation frequency. The top middle plot of figure \ref{frequencies} shows the relative difference of the three analytic orbital frequencies with respect to the numerical orbital frequency, while the top right and the bottom left show the relative differences of radial and the vertical oscillation frequencies respectively. One can see that the approximate spacetime captures the behaviour of the orbital frequency quite accurately, almost 1 order of magnitude better than the other two spacetimes, and it does quite well with respect to the other two frequencies as well.   

The frequencies are important physical characteristics of the spacetime and have direct astrophysical relevance since they can be related to QPOs. In particular it is important for the analytic spacetime to be able to capture accurately a behaviour of the frequencies that is characteristic of NSs and is absent in the case of black holes. That is the behaviour of the nodal precession frequency $\Omega_z=\Omega-\kappa_z$, which for NSs can display a turnover at inner radii and in some cases can become even negative, as it has been discussed in the past by \cite{Morsink1999ApJ,PappasQPOs,Gondeketal,Pappas2015Unified,Tsang-Pappas2016ApJ}. The turnover effect is demonstrated at the bottom middle plot in figure \ref{frequencies}, where one can see that the proposed approximate spacetime exhibits the turnover behaviour in the same way as the numerical spacetime does, while the other two metrics fall either too far or too short of the right behaviour. Here we will briefly sketch why this happens, but a detailed investigation will appear elsewhere. 

The reason why the approximate metric does better than the other two metrics and traces the numerical frequency that well is no surprise and has to do with the contribution of each multipole moment to the nodal precession frequency. If one expresses $\Omega_z$ as an expansion in terms of inverse powers of $R_{circ}$ and multipole moments, one can separate the contribution that the individual multipole moments have. Then one can see that the mass quadrupole and the mass hexadecapole have negative contributions, while the angular momentum and the spin octupole have positive contributions. This means that the Hartle\&Thorne spacetime, which has vanishing spin octupole and mass hexadecapole \citep[see][for details]{twosoliton}, is missing a positive contribution from the $S_3$ and the negative quadrupole contribution turns the frequency over early with respect the other curves. On the other hand, the two-soliton spacetime has the right $S_3$ contribution but underestimates $M_4$, which for this spacetime depends on the values of the first four moments. This means that the frequency is missing a negative contribution from $M_4$ that would have contributed further in (higher order moments contribute more at smaller radii) missing in this way the turnover. Finally, the approximate spacetime has the correct moments contributions up to the $M_4$ and this leads to the frequency turning over in the same way as the numerical spacetime.    

This behaviour, i.e., the turning over of the nodal precession frequency, is an effect that is absent in the Kerr geometry where the higher order moments have $\alpha=\beta=\gamma=\ldots=1$. As we have discussed previously for the effect to manifest one needs a spacetime with higher order mass moments that are larger than that of Kerr, which in the case of NS spacetimes translates to having $\alpha\go4$ (the plot in figure \ref{frequencies} corresponds to $\alpha=4.209$) depending on the rotation rate (spin parameter $j$). The models that have large enough $\alpha$ are the low-mass models with masses around and lower than the ``canonical'' mass of $1.4M_{\odot}$, depending on the EoS. Recent studies have shown that such objects exist in the range around $1.25M_{\odot}$ as parts of X-ray binaries \citep[see]{Valentim2011MNRAS,Valentim2016arXiv}, therefore this behaviour of the nodal precession frequency might have some interesting phenomenology in these systems \citep[see][for example]{Tsang-Pappas2016ApJ} and it will be important for NS spacetime models to be able to capture it.

The last property that we compare is the energy change per logarithmic change in the orbital frequency, $\Delta\tilde{E}$. The relative difference with respect to the numerical spacetime for the proposed approximate metric and the other two metrics is show in the bottom right plot of figure \ref{frequencies}. One can see that the proposed metric does well and specifically it does much better than the other two.

In general, the comparison has shown that the proposed approximate metric captures very well overall and individually the various properties and characteristics of a NS spacetime and can therefore be used to describe the exterior of NSs.

\section{The geometry around scalarized NSs in scalar-tensor theory}
\label{sec:conclusions}

Up to this point we have discussed NSs in the context of GR. Compact objects in general and NSs in particular are of interest with respect to theoretical extensions of GR.  
Scalar-tensor theory is one of the most extensively studied extensions of GR \citep{Jordan49,fierz56,Jordan59,BransDicke,Dicke62,Damour,FujiiMaeda,CapozzielloFaraoni} and there has been a lot of work on studying NSs in this theory \citep[see][for example]{Damour93PRL,Horbatsch2011JCAP,Doneva2013PhRvD,DonevaQPOs2014PhRvD,Doneva2014PhRvD}. The theory can be derived by the following action
\begin{equation}
\label{staction}
S=\int d^4x \sqrt{-g} \left(\Phi R-\frac{\omega(\Phi)}{\Phi} \nabla^\mu \Phi \nabla_\mu \Phi\right)+S_m(g_{\mu\nu},\psi)\,,
\end{equation}
where $g$ is the determinant and $R$ is the  Ricci scalar of the metric $g_{\mu\nu}$, $\nabla_\mu$ denotes the corresponding covariant  derivative, $S_m$ is the matter Lagrangian, and $\psi$  collectively denotes the matter fields which are taken to couple minimally to the metric $g_{\mu\nu}$ (not to be confused with the scalar twist introduced in section \ref{sec:2}). 

Respectively the scalar field $\Phi$ is nonminimally coupled to gravity and has a noncanonical kinetic term. This representation of the theory is called the Jordan frame (or physical frame). The conformal transformation $\tilde{g}_{\mu\nu}=16 \pi G\, \Phi \, g_{\mu\nu}$, together with the scalar field redefinition
\be
\label{scalarredef}
d\phi=\sqrt{\frac{2\omega(\Phi)+3}{4}} \, d\ln\Phi\,,
\ee
bring action (\ref{staction}) to the following form
\begin{equation}
\label{stactionein}
S=\frac{1}{16\pi G}\int d^4x \sqrt{-\tilde{g}} \left(\tilde{R}-2 \tilde{\nabla}^\mu \phi \tilde{\nabla}_\mu \phi\right)+S_m(g_{\mu\nu},\psi)\,,
\end{equation}
where the matter fields still couple minimally to $g_{\mu\nu}$, but the redefined scalar field $\phi$ is now coupled to the matter fields, and it is this coupling that encodes any deviation from standard GR with a minimally coupled scalar field. This frame is called the Einstein frame. The advantage of the Einstein frame is that the field equations outside the matter sources take the form,
\begin{align}
\label{feq1}
\tilde{R}_{ab} &= 2 \partial_a\phi\partial_b\phi,\\
\tilde{g}^{ab}\tilde{\nabla}_a\tilde{\nabla}_b\phi &=0,
\end{align}  
which are essentially GR with a minimally coupled scalar field. 

This similarity to GR allows \citep[see for more details][]{PappasSTMoments,PappasSTRyan,Cardoso2016CQG}, in the stationary and axisymmetric cases, for the possibility of casting the field equations of scalar-tensor theory in the form of an Ernst equation (\ref{ErnstE}), as in GR, with the addition of a Laplace equation for the scalar field, 
\be \nabla^2 \phi=0\ee
where, as for the Ernst equation, $\nabla$ is the gradient in 3-dimensional flat cylindrical coordinates, while the line element has the Papapetrou form (\ref{Pap}) and the metric function $\gamma(\rho,z)$ is given in this case by the modified equations, 
\be \frac{\p\gamma}{\p \rho}=\left(\frac{\p\gamma}{\p \rho}\right)_{GR}+\rho \left[\left(\frac{\p\phi}{\p \rho}\right)^2-\left(\frac{\p\phi}{\p z}\right)^2\right], \label{gamma1}\ee
\be \label{gamma2} 
\frac{\p\gamma}{\p z}=\left(\frac{\p\gamma}{\p z}\right)_{GR}+2 \rho \left(\frac{\p\phi}{\p \rho}\right)\left(\frac{\p\phi}{\p z}\right).\ee
 
This means that with respect to the Ernst equation and the metric functions $f(\rho,z)$ and $\omega(\rho,z)$, the equations will be exactly the same as before and a solution can be constructed in the same way as in GR in terms of the potential $\xi$ given in equations (\ref{xi},\ref{xi2}). The new component that is introduced and modifies the equations for the function $\gamma(\rho,z)$, i.e., the scalar field $\phi$, will have to be expressed, as $\xi$ was, in terms of an asymptotic expansion,
\be \tilde{\phi}=(1/\bar{r})\phi=  \sum_{i,j=0}^{\infty} b_{ij}\bar{\rho}^i\bar{z}^j,\ee
where, from the Laplace equation that $\phi$ must satisfy, we have that the coefficients $b_{ij}$ should satisfy the recursive relation, 
\be b_{i+2,j}=-\frac{(j+2)(j+1)}{(i+2)^2} b_{i,j+2}, \label{bCoeff}\ee
and that $b_{1,j}=0$. If one further assumes reflection symmetry about the equatorial plane, then the  coefficients of odd powers of $\bar{z}$ should be $b_{i,2j+1}=0$. As was the case for $\xi$, all the $b_{ij}$ coefficients can be calculated from the expansion of $\tilde{\phi}$ along the symmetry axis, $\tilde{\phi}(\bar{\rho}=0)=\sum_{j=0}^{\infty} w_j\bar{z}^j$ \citep[see][]{PappasSTMoments}. The multipole moments in scalar-tensor theory are expressed in terms of the $w_j$ and $m_j$ coefficients, therefore $w_j$ and $m_j$ can be expressed in terms of the scalar moments $W_n$ and the mass, $M_n$, and angular momentum $S_n$, moments.   

Proceeding to write down the metric functions $f(\rho,z)$, $\omega(\rho,z)$, and $\gamma(\rho,z)$ in terms of the scalar-tensor multipole moments we have, 

\bea  f(\rho,z) \!\!\!\!\!\!&=&\!\!\!\!\!\!1-\frac{2 M}{\sqrt{\rho ^2+z^2}}+\frac{2 M^2}{\rho ^2+z^2}+\frac{ C^{ST}(\rho,z)}{3 \left(\rho ^2+z^2\right)^{5/2}} \nn\\
                      &&\!\!\!\!\!\! +\frac{ D^{ST}(\rho,z)}{3 \left(\rho^2+z^2\right)^3} +\frac{A^{ST}(\rho,z) }{420 \left(\rho ^2+z^2\right)^{9/2}}\nn\\
                       &&\!\!\!\!\!\! +\frac{B^{ST}(\rho,z) }{630 \left(\rho ^2+z^2\right)^5},\\
           \omega(\rho,z) \!\!\!\!\!\!&=&\!\!\!\!\!\! -\frac{2 J \rho ^2}{\left(\rho ^2+z^2\right)^{3/2}}-\frac{2 J M \rho ^2}{\left(\rho ^2+z^2\right)^2}  +\frac{ F^{ST}(\rho,z)}{5 \left(\rho ^2+z^2\right)^{7/2}}\nn\\
                                        &&\!\!\!\!\!\!   +\frac{H^{ST}(\rho,z) }{30 \left(\rho ^2+z^2\right)^4}  + \frac{ G^{ST}(\rho,z)}{60 \left(\rho ^2+z^2\right)^{11/2}}  ,   \\
            \gamma(\rho,z) \!\!\!\!\!\!&=&\!\!\!\!\!\!   \frac{\rho ^2}{4 \left(\rho ^2+z^2\right)^4}  \left[ \rho ^2 \left(J^2+M^4\right)-4 z^2 \left(2 J^2+M^4\right) \right.\nn\\
                                         &&\!\!\!\!\!\!\!\!  \left.  -\left(W_0 \left(2 M^2 W_0+W_0^3+3 W_2\right)+3 M M_2\right) (4 z^2-\rho^2 )\right]  \nn\\
                                     &&\!\!\!\!\!\!      -\frac{\rho ^2 \left(M^2+W_0^2\right)}{2 \left(\rho ^2+z^2\right)^2}   ,                                  
   \eea
where,

\bea   C^{ST}(\rho,z) \!\!\!\!\!\!&=&\!\!\!\!\!\! \left[\rho ^2 \left(3(M_2- M^3)+M W_0^2\right)\right.\nn\\
                                                  &&\left.-2 z^2 \left(3 \left(M^3+M_2\right)+M W_0^2\right)\right], \\
          D^{ST}(\rho,z) \!\!\!\!\!\!&=&\!\!\!\!\!\! \left[ 2 z^2 \left(M \left(3 M^3+2 M W_0^2+6 M_2\right)-3 J^2\right)\right.\nn\\
                                               &&\left.-2 M \rho ^2 \left(M W_0^2+3 M_2\right)\right] ,
                                               \eea
\bea               A^{ST}(\rho,z) \!\!\!\!\!\!&=&\!\!\!\!\!\!   \left[8 \rho ^2 z^2 \left(360 J^2 M+91 M^3 W_0^2+255 M^2 M_2\right.\right.\nn\\
                                   &&\!\!\!\!\!\!       \left. +63 M W_0^4+270 M_2 W_0^2+90 M W_2 W_0+315  M_4\right)\nn\\
                                    &&\!\!\!\!\!\!      -\rho ^4 \left(150 J^2 M-105 M^5-154 M^3 W_0^2 \right. \nn\\
                                    &&\!\!\!\!\!\! -480 M_2 M^2+63 M W_0^4+90 M W_0 W_2\nn\\
                                     &&\!\!\!\!\!\!  \left.+270 M_2 W_0^2+315 M_4\right)-8 z^4 \left(-300 J^2 M \right.\nn\\
                                      &&\!\!\!\!\!\! +105 M^5+112 M^3 W_0^2+330 M_2 M^2+21 M W_0^4\nn\\
                                       &&\!\!\!\!\!\!   \left.\left.+30 M W_0 W_2+90 M_2 W_0^2+105M_4\right)  \right],   \eea
\bea         
    B^{ST}(\rho,z) \!\!\!\!\!\!&=&\!\!\!\!\!\! \left[ \rho ^4 \left(M \left(2 M \left(225 J^2+84 M^2 W_0^2+112   W_0^4\right.\right.\right.\right.\nn\\
                                  &&\!\!\!\!\!\!\!\!\! \left.\left.+135 W_2 W_0\right)+945 M_4\right)+30 M_2 \left(15 M^3 \right.\nn\\
                                 &&\!\!\!\!\!\!  \left.\left.+34 M W_0^2\right)+315 M_2^2\right)\nn\\
                                    &&\!\!\!\!\!\!  +4 z^4 \left(-18 \left(J \left(100 J M^2+21 J W_0\right.\right.\right.\nn\\
                                &&\!\!\!\!\!\!   \left.\left.+35 S_3\right)-35 M M_4\right)+150 M_2 \left(9 M^3+5 M W_0^2\right) \nn\\
                                 &&\!\!\!\!\!\!  +M^2 \left(315 M^4+462 M^2  W_0^2+161 W_0^4\right.\nn\\
                                      &&\!\!\!\!\!\!  \left.  \left.  +180 W_0 W_2\right)+315 M_2^2\right)\nn\\
                                  &&\!\!\!\!\!\! -4 \rho ^2 z^2 \left(27 \left(J \left(45 J M^2-21 J W_0 \right.\right.\right.\nn\\
                              &&\!\!\!\!\!\!   \left.\left.   -35S_3\right)+70 M M_4\right)+30 M_2 \left(72 M^3+61 M W_0^2\right)\nn\\
                               &&\!\!\!\!\!\!   +M^2 \left(315 M^4+756 M^2 W_0^2+413 W_0^4 \right.\nn\\
                                 &&\!\!\!\!\!\!  \left. \left. \left. +540 W_0  W_2\right)   +315 M_2^2\right)\right], \eea
\bea 
  H^{ST}(\rho,z) \!\!\!\!\!\!&=&\!\!\!\!\!\!    \left[\rho ^2 \left(M \left(-120 J M^2 z^2+J W_0 \left(5 W_0 \left(\rho ^2+4 z^2\right)\right.\right.\right.\right.\nn\\
                                        &&\!\!\!\!\!\!   \left.\left.      +27 \left(\rho ^2-4 z^2\right)\right)+45 S_3 \left(\rho  ^2-4 z^2\right)\right)  \nn\\
                                        &&\!\!\!\!\!\! \left.\left.+15 J M_2 \left(\rho ^2+4 z^2\right)\right)\right],\\
            G^{ST}(\rho,z) \!\!\!\!\!\!&=&\!\!\!\!\!\!    \left[\rho ^2 \left(15 J \left(\rho ^4 \left(M^4-J^2\right)-8 z^4 \left(J^2+3 M^4\right) \right.\right.\right. \nn\\
                                         &&\!\!\!\!\!\! \left. +4 \rho ^2 z^2 \left(3 J^2+M^4\right)\right)+M^2  \left(J W_0 \left(10 W_0 \left(\rho ^4 \right.\right.\right. \nn\\
                                         &&\!\!\!\!\!\! \left.\left.-8 z^4+20 \rho ^2 z^2\right)+9 \left(3 \rho ^4-40 z^4+12 \rho ^2 z^2\right)\right)   \nn\\
                                         &&\!\!\!\!\!\!   \left.+15 S_3 \left(3  \rho ^4-40 z^4+12 \rho ^2 z^2\right)\right)\nn\\
                                         &&\!\!\!\!\!\! \left.\left. +30 J M_2 M \left(\rho ^4-8 z^4+20 \rho ^2 z^2\right)\right)\right],\\
         F^{ST}(\rho,z) \!\!\!\!\!\!&=&\!\!\!\!\!\! \left[ \rho ^2 \left(-5 J M^2 \left(\rho ^2+4 z^2\right) \right.\right.\nn\\
                                         &=&\!\!\!\!\!\! \left.\left.-\left(4 z^2-\rho ^2\right) \left(3 J W_0+5 S_3\right)\right)\right].
   \eea
These metric functions, together with the scalar field, 
\be \phi(\rho,z)=\frac{W_0}{\sqrt{\rho^2+z^2}}\left[1-\frac{\left(M^2 W_0+W_0^3+3 W_2\right) \left(r^2-2 z^2\right)}{6 W_0 \left(r^2+z^2\right)^2}\right] \ee
 constitute a solution of the Einstein field equations, $\tilde{g}_{\mu\nu}$, and the scalar field equation in the Einstein frame. For the calculation of the metric functions and the scalar field again we have assumed as in GR an expansion which is truncated at order $\mathcal{O}(\bar{r}^6)$. This solution can approximate up to the given order any spacetime that is characterised be a given set of moments. If the moments are chosen so as to correspond to a scalarized NS, then the spacetime will correspond to that of the particular NS. 
 
 To connect this spacetime to observables, one should perform a conformal transformation and go to the physical or Jordan frame, since particles follow the geodesics of the metric of that frame. One can have a general description of the spacetime in the Jordan frame if one assumes a conformal factor of the form $\Phi^{-1}=A^2(\phi)$, which can then be Taylor expanded around the value of the scalar field at infinity \citep[see][]{PappasSTRyan}. This expansion therefore can be written in terms of the coupling parameters defined by \cite{Damour,Damour96PRD} and therefore one can have a general description of the spacetime without subscribing to a particular choice for the conformal factor. The Jordan frame metric will then be, 
 \be g_{\mu\nu}=A^2(\phi)\tilde{g}_{\mu\nu}. \ee

\section{Conclusions}
\label{sec:conclusions}

In this work we have presented an approximate solution for the spacetime around NSs that is parametrised by the first multipole moments up to the mass hexadecapole, i.e., the moments $M$, $J=S_1$, $Q=M_2$, $S_3$, and $M_4$. This solution is produced using the \cite{ernst1} formulation of GR. 

Furthermore, we have taken advantage of the recently discovered 3-hair relations for the NS multipole moments \citep{Pappas:2013naa,YagietalM4} to present an EoS independent description of the NS spacetime that depends on only three parameters, the mass $M$, the spin parameter $j=J/M^2$, and the quadrupolar deformability $\alpha=-M_2/(j^2 M^3)$. 

The resulting approximate spacetime has been compared against numerically constructed NS spacetimes for different realistic EoSs and has been shown to be of better accuracy than previously tested analytic spacetimes, such as the two-soliton spacetime or the exterior Hartle~\&~Thorne spacetime, in the range $1.5-2\lo\alpha\lo10$. In particular the most interesting characteristic of this new analytic spacetime is the more accurate description of geodesic properties such as the orbital and precession frequencies of equatorial orbits. This is of particular astrophysical relevance since these frequencies can be associated to observables such as QPOs. An interesting result of the analysis presented here is that the mass hexadecapole $M_4$ plays a part in shaping the profile of the nodal precession frequency in the region close to the innermost stable circular orbit even for a moderate quadrupolar deformability ($\alpha\simeq4.2$). A detailed analysis of this is a work in progress. 

Additionally to the accuracy, the advantage of the proposed approximate spacetime is its functional simplicity \citep[the metric is no more complicated than the exterior Hartle~\&~Thorne metric, see][for example]{berti-white} which makes it more attractive to use than other analytic solutions that have been proposed so far \citep{Wattsetal2016RvMP}. In addition, the parameterisation of the spacetime with respect to the mass, the spin and the quadrupole moment could prove to be useful for solving the inverse problem of determining the EoS from observations \citep[see][]{Pappas:2013naa,Pappas2015Unified}. 

Finally, the approximate spacetime has been extended to the scalar-tensor theory of gravity with a massless scalar field. What remains to be seen is whether the scalarised spacetime will be as accurate as its GR counterpart in describing the exterior of NSs. This can be done by calculating NS models and the corresponding spacetime in these theories \citep[see work by][on developing the numerical codes for rapidly rotating NSs in scalar-tensor theory]{Doneva2013PhRvD,DonevaQPOs2014PhRvD,Doneva2014PhRvD} and using them to test the approximate analytic spacetime presented here.

As a final note we should mention that, if needed, the approximate spacetime presented here can be easily calculated at a higher order of accuracy by taking into account the contribution of higher order moments and keeping higher order terms in the expansion. This can be done without a significant increase in the complexity of the final analytic spacetime.

\section*{Acknowledgments}

G.P. would like to thank Kostas Glampedakis, Hector O. Silva and E. Berti for their useful comments and suggestions on early versions of the manuscript.
G.P. was supported by NSF CAREER Grant No. PHY-1055103. This work was supported by the H2020-MSCA-RISE-2015 Grant No. StronGrHEP-690904 and FCT contract IF/00797/2014/CP1214/CT0012 under the IF2014 Programme.


\bibliographystyle{mnras} 
\bibliography{mn-jour,mybibliography}

\appendix
%
\section[]{Coefficients $a_{ij}$ in GR and in Scalar-Tensor theory}
\label{app:A}

In the main text the metric in both GR and scalar-tensor theory is constructed by an ansatz for the potential $\xi$ that is given in equation (\ref{xi2}) in terms of the coefficients $a_{ij}$. Here we give the relation of these coefficients to the $m_i$ coefficients, which are related to the multipole moments, 

\begin{align}
                     a_{20} &=-\frac{1}{2} \left(m_0^* m_0^2+m_2\right), \\
                     a_{21} &=-\frac{1}{2} \left(3 m_3+m_0 \left(4 m_1 m_0^*+m_0 m_1^*\right)\right),\\
                     a_{22} &=\frac{1}{2} \left(-4 m_0^* m_1^2-4 m_0 m_1^* m_1-6 m_4\right.\nn\\
                                & \left.-m_0 \left(6 m_2  m_0^*+m_0 \left(m_0 \left(m_0^*\right)^2+m_2^*\right)\right)\right),\\
                     a_{23} &= -2 m_1^* m_1^2-\frac{1}{2} \left(5 m_0^2 \left(m_0^*\right)^2+12 m_2
   m_0^*+4 m_0 m_2^*\right) m_1\nn\\
                                 &-5 m_5-\frac{1}{2} m_0 \left(6 m_2   m_1^*+2 m_0^* \left(m_1^* m_0^2+4 m_3\right)+m_0 m_3^*\right) ,\\
                     a_{40} &=  \frac{1}{8} \left(-m_0^* m_1^2+m_0 \left(2 m_0^2 \left(m_0^*\right)^2+3 m_2
   m_0^* \right.\right.   \nn\\
                                 &\left.-\frac{1}{2} m_0 \left(-m_0
   \left(m_0^*\right)^2-m_2^*\right)\right)+\frac{1}{2} \left(4 m_0^* m_1^2+4 m_0 m_1^* m_1\right. \nn\\
                                 &\left.\left.+6 m_4+m_0 \left(6 m_2 m_0^*+m_0 \left(m_0  \left(m_0^*\right)^2+m_2^*\right)\right)\right)\right) , \\
                    a_{41} &=  \frac{1}{8} \left(5 m_0^* m_1^* m_0^3+\left(15 m_1 \left(m_0^*\right)^2+\frac{1}{2} \left(m_1 \left(m_0^*\right)^2\right.\right.\right. \nn\\
                                &\left.\left.+4 m_0  m_1^* m_0^*+3 m_3^*\right)\right) m_0^2+\left(12 m_3  m_0^*+3 m_2 m_1^*\right.\nn\\
                                &\left.-2 m_1 \left(-m_0 \left(m_0^*\right)^2-m_2^*\right)\right) m_0-m_1^2 m_1^*-3 \left(-2 m_1^* m_1^2 \right.\nn\\
                                &-\frac{1}{2} \left(12 m_2 m_0^*+m_0 \left(5 m_0
   \left(m_0^*\right)^2+4 m_2^*\right)\right) m_1-5 m_5 \nn\\
                                &\left.\left.-\frac{1}{2} m_0 \left(8 m_3
   m_0^*+6 m_2 m_1^*+m_0 \left(2 m_0 m_0^* m_1^*+m_3^*\right)\right)\right)\right).
\end{align}
The relations between these $m_i$ coefficients and the moments can be found in the work by \cite{fodor:2252,PappasSTMoments} for GR and scalar-tensor theory respectively. For the case of scalar-tensor theory one also needs the $b_{ij}$ coefficients in terms of the $w_j=b_{0j}$ coefficients. These are calculated from equation (\ref{bCoeff}).

\section[]{$\bar{M}_4-\bar{M}_2$ fit for NSs}
\label{app:B} 
 
\begin{figure}
\centering
\includegraphics[width=.6\textwidth]{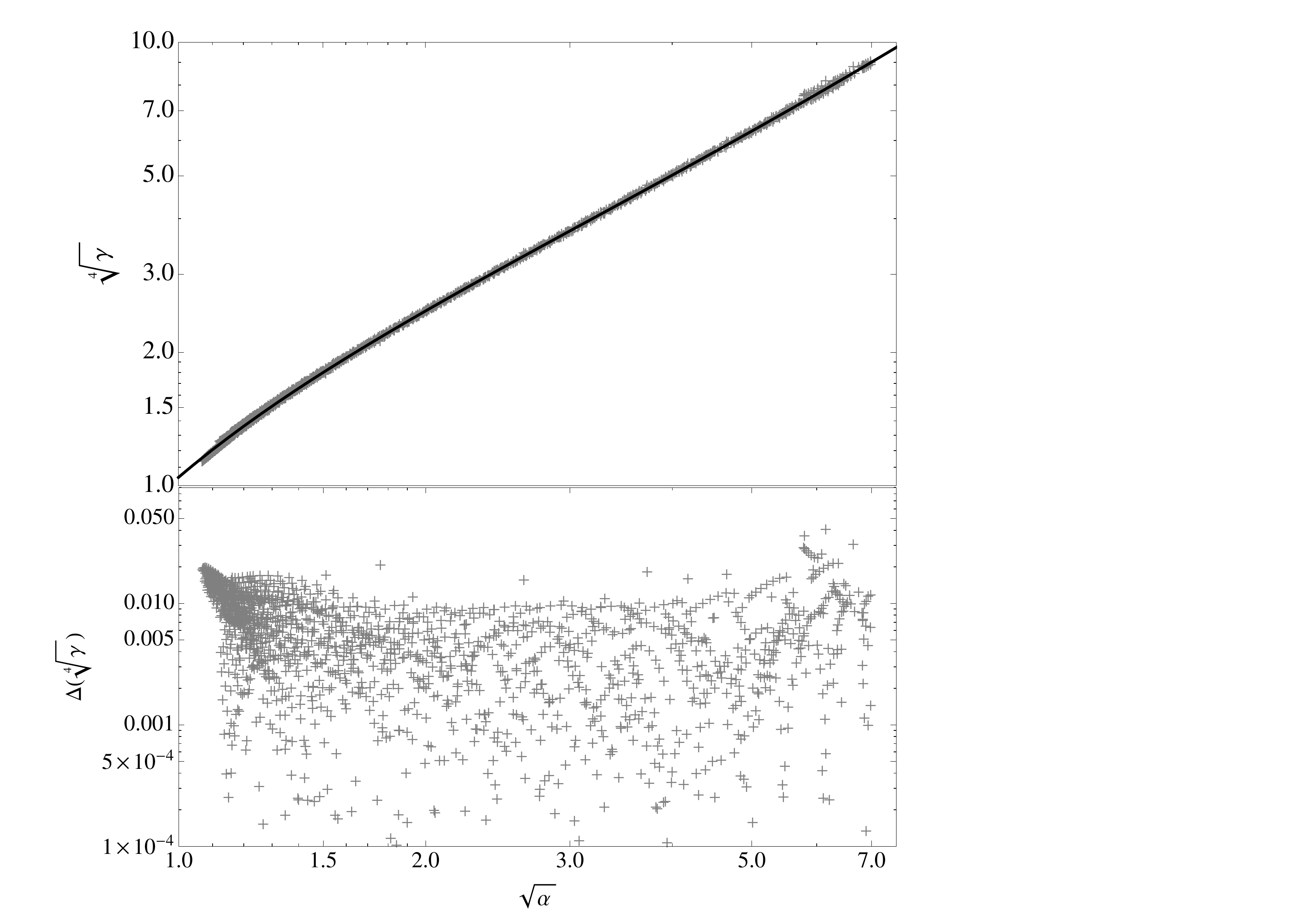}
\caption{Plot of $\sqrt[4]{\bar{M}_4}$ as a function of $\sqrt{-\bar{M}_2}$ for various realistic EoSs (top plot). All the models constructed fall on the same line. The accuracy of the fitting formula is better than 5 per cent (bottom plot).}
\protect\label{M4fit}
\end{figure}

 Using only the NS data of \cite{YagietalM4} one can try and fit the mass hexadecapole as a function of the mass quadrupole using a fitting expression of the form of Eq. (\ref{momentsFit}). The result of that fit is the expression presented in the main text, i.e.,
 \be  y_2=-4.749+0.27613\, x^{1.5146}+5.5168\, x^{0.22229},\ee
where $y_2=\sqrt[4]{\bar{M}_4}=\sqrt[4]{\gamma}$ and $x=\sqrt{-\bar{M}_2}=\sqrt{\alpha}$. The accuracy of this fit is shown in Figure \ref{M4fit}, where on the upper panel we have plotted the NS data points (reduced $M_4$ against reduced $M_2$) using grey crosses with the best fit curve going through them, while the bottom panel shows the relative difference of the fit from the actual NS values. One can see that the relative difference is always better than $4-5$ per cent. The fact that one could achieve better accuracy in the fitting relation between the moments if one considered only NSs had already been noted by \cite{YagietalM4}. Figure \ref{M4fit} is equivalent to Figure 1 by \cite{YagietalM4}.
 
\section[]{Relation of $\tilde{\beta}_0$ and $\tilde{\beta}_2$ to $\bar{M}_2$ for NSs}
\label{app:C} 

It was mentioned in section \ref{sec:QuasIsoCoord} that the coefficients $\tilde{\beta}_0$ and $\tilde{\beta}_2$, of the $B$ function expansion in the quasi-isotropic coordinates, could be expressed in terms of $\alpha$ in an approximately EoS independent way. For completeness we present in figure \ref{fig:b0b2} the relation of these coefficients to $\alpha$ for the same models and EoSs used in Appendix \ref{app:B}.

\begin{figure}
\centering
\includegraphics[width=.45\textwidth]{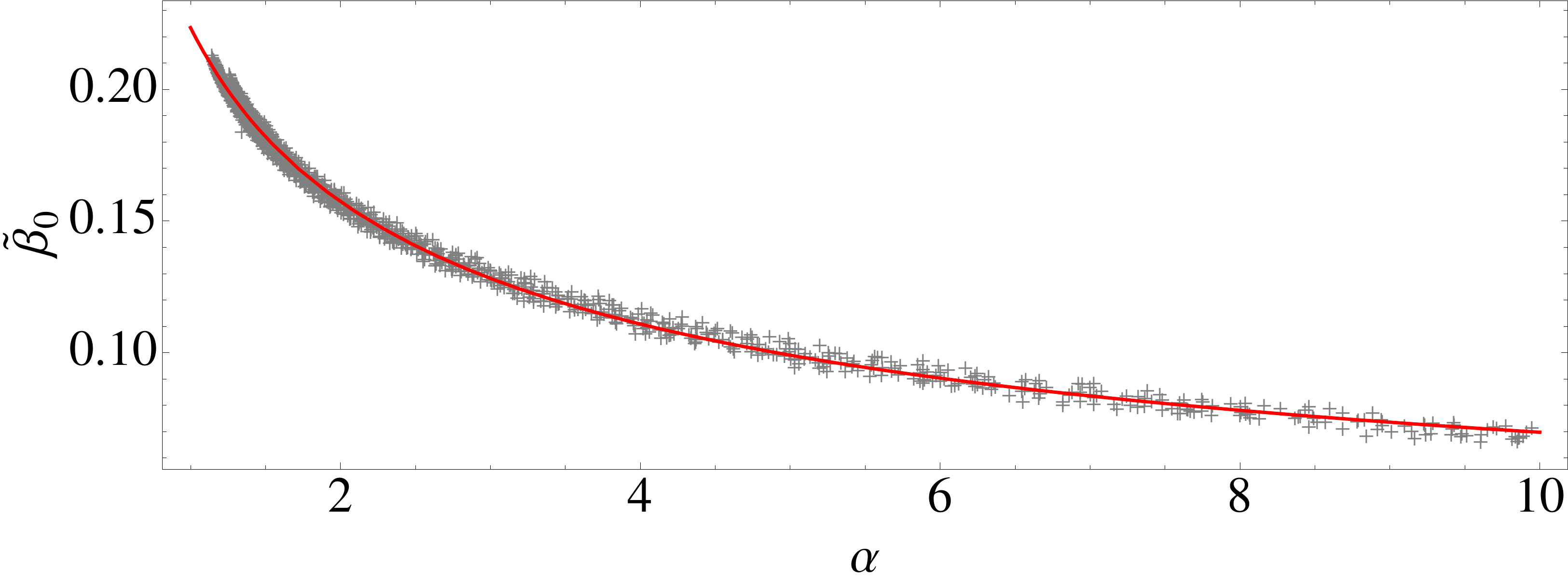}
\includegraphics[width=.44\textwidth]{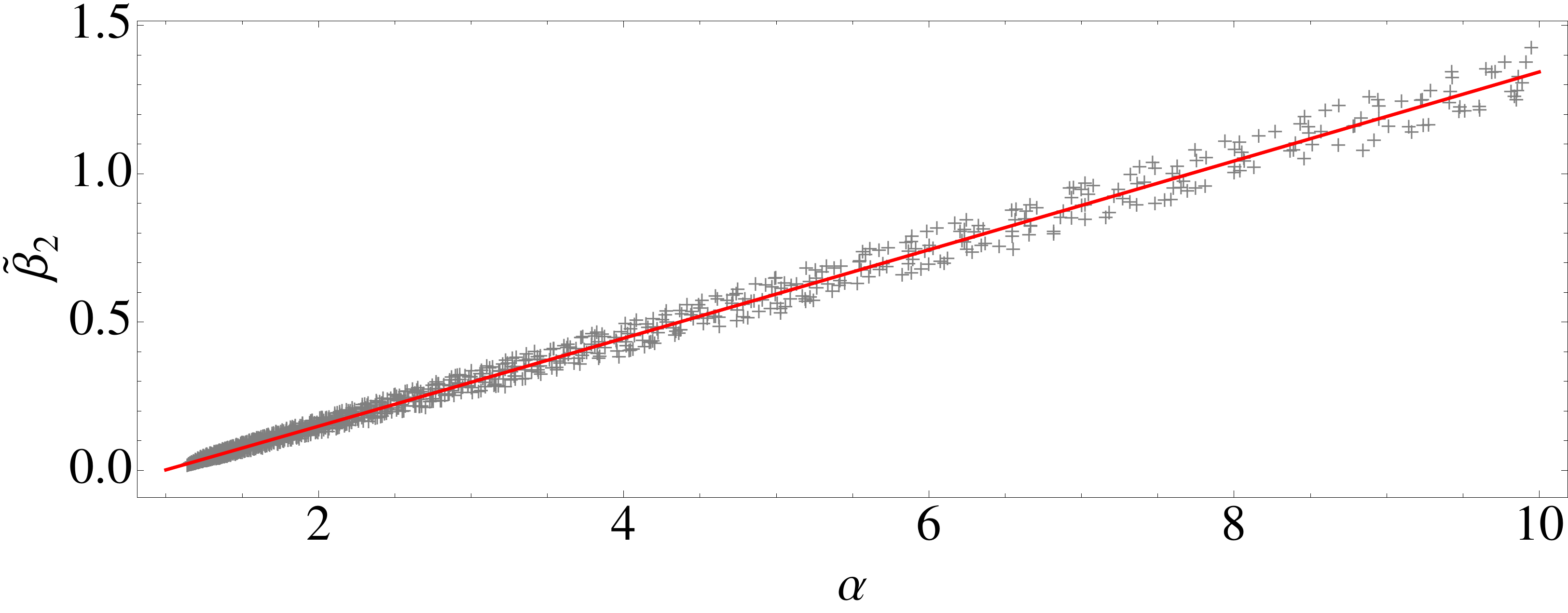}
\caption{Plots of $\tilde{\beta}_0$ (upper) and $\tilde{\beta}_2$ (lower) as a function of $-\bar{M}_2\equiv\alpha$ for various realistic EoSs.  The data could be fitted with a curve of the form $y=A+Bx^{\nu}$.}
\protect\label{fig:b0b2}
\end{figure}

\bsp \label{lastpage}

\end{document}